\documentclass[iop,trackchanges,linenumbers]{emulateapj}
\usepackage{graphicx}
\usepackage{apjfonts}
\usepackage{amsmath} 
\usepackage{natbib}
\usepackage{color}
\usepackage[normalem]{ulem}
\DeclareOldFontCommand{\rm}{\normalfont\rmfamily}{\mathrm}
\newcommand{\prima}{^{\prime}}
\newcommand{\primast}{^{\prime\ast}}
\newcommand{\fdeg}{.\!\! ^{\circ}}
\newcommand{\nprimac}{n^{\prime 2}}
\def\Etilde{\tilde{\textbf {E}}^{(\text{t})}}
\def\Htilde{\tilde{\rm \textbf{H}}}
\def\Htildeel{\tilde{\rm H}}
\def\Htildenb{\tilde{\rm H}}
\def\Et{\textbf {E}^{(\text{t})}}
\def\Ei{\textbf {E}^{(\text{i})}}
\def\E{\textbf {E}}

\def\ee{\hat{\textbf{e}}_3}
\def\s{\hat{\textbf{s}}}

\def\H{{\rm H}}

\def\Iin{I^{\rm (i)}}
\def\Qin{Q^{\rm (i)}}
\def\Airy{\left(\frac{2J_1(z)}{z}\right)^2}

\def\degree{^{\circ}}
\def\bepsilon{\boldsymbol{\epsilon}}
\def\imperm{{\rm \boldsymbol{\eta}}}
\def\Hmatrix{{\rm \textbf{H}}}
\def\Mmatrix{{\rm \textbf{M}}}
\def\Mijtilde{{\rm \tilde{M}\prima}}
\def\PSF{{\cal S}}
\def\taueff{\tau_{\rm eff}}
\def\tauprima{\tau_{\rm eff}^\prime}
\definecolor{al}{rgb}{0.9,0.1,0.0}

\slugcomment{Submitted to ApJS}

\shorttitle{The anisotropic (birefringent) case}
\shortauthors{F.J. Bail\'en et al.}

\begin{document}

\title{On Fabry-P\'erot etalon based instruments \\
	II. The anisotropic (birefringent) case}

\author{F.J. Bail\'en, D. Orozco Su\'arez, and J.C. del Toro Iniesta}
\affil{Instituto de Astrof\'isica de Andaluc\'ia (CSIC), Apdo. de Correos 3004, E-18080 Granada, Spain}
\email{fbailen@iaa.es,orozco@iaa.es,jti@iaa.es}

\begin{abstract}
Crystalline etalons present several advantages with respect to other types of filtergraphs when employed in magnetographs. Specially that they can be tuned by only applying electric fields. However, anisotropic crystalline etalons can also introduce undesired birefringent effects that corrupt the polarization of the incoming light. In particular, uniaxial Fabry-P\'erots, such as LiNbO$_3$ etalons, are birefringent when illuminated with an oblique beam. The farther the incidence from the normal, the larger the induced retardance between the two orthogonal polarization states. The application of high-voltages, as well as fabrication defects, can also change the direction of the optical axis of the crystal, introducing birefringence even at normal illumination. Here we obtain analytical expressions for the induced retardance and for the Mueller matrix of uniaxial etalons located  in both collimated and telecentric configurations. We also evaluate the polarimetric behavior of $Z$-cut crystalline etalons with the incident angle, with the orientation of the optical axis, and with the f-number of the incident beam for the telecentric case. We study artificial signals produced in the output Stokes vector in the two configurations. Last, we discuss the polarimetric dependence of the imaging response of the etalon for both collimated and telecentric setups.
\end{abstract}


\section{Introduction} \label{sec:intro}
Narrow-band tunable filters are widely used in solar physics to carry out high precision imaging in selected wavelength samples. In the particular case of Fabry-P\'erot etalons, the sampling can be done by either modifying the refraction index of the material, by changing the width of the Fabry-P\' erot cavity or both. Naturally, temperature fluctuations and variations of the tilt angle of the etalon plates with respect to the incident light change their tunability as well.

The more common technology used in Fabry-P\'erots in ground-based instruments is that of piezo-stabilized etalons \citep[e.g.,][]{kentischer,ref:puschman,ref:scharmer}. For space applications, however, they are very demanding in terms of total weight or mounting, to name a few. Solid etalons based on electro-optical and piezo-electric material crystals are way lighter and do not need the use of piezo-electric actuators, thus avoiding the introduction of mechanical vibrations in the system. Examples are LiNbO$_3$ or MgF$_2$ based etalons and liquid crystal etalons \citep[e.g.,][]{ref:alvarez,ref:gary}, which can be tuned after modification of a feeding voltage signal.
 Crystals used in these Fabry-P\'erot etalons are typically birefringent and therefore able to modify the polarization of light. The risk for uncertainties in the measured Stokes parameters, hence altering the polarimetric efficiencies of the system is not null and should be assessed. 

In liquid crystal etalons, the optical axis direction depends on the electric field applied and, therefore, birefringence will not only change with the incident direction, but also when tuning the etalon. To avoid this effect, lithium niobate or magnesium fluoride etalons can be used with given cut configurations that select their constant optical axis. Etalons with the optical axis parallel to the reflecting surfaces ($Y$-cut) are used sometimes \citep[e.g.,][]{ref:netterfield} but the $Z$-cut configuration is often preferred \citep[][]{imax,ref:sophi}, since the optical axis is perpendicular to the reflecting surfaces of the etalon and, as a result, no polarization effects are expected for normal illumination. Although close to normal, typical instruments receive light from a finite aperture. Hence, spurious polarization effects cannot be neglected without an analysis. Moreover, local inhomogeneities of the crystals and other fabrication defects can modify the crystalline (birefringent) properties of the etalons.

 Most efforts have been driven, so far, to study the propagation of the ordinary and extraordinary ray separately in some particular cases. One example is the work by \cite{ref:doerr}, where spurious polarization effects due to oblique illumination in Fabry-P\'erots have been studied numerically by considering the influence of thin film multilayer coatings in isotropic etalons. Another example can be found in \cite{ref:vogel}, where experimental results on the polarization-dependent transmission in liquid (uniaxial) crystals are presented.
On the other hand, \cite{ref:jti} modeled the polarimetric response of uniaxial etalons as retarders to include their effect on the polarimetric efficiency of modern magnetographs and \cite{ref:lites} obtained an analytical expression for the Mueller matrix of a linear retarder (i.e., a crystalline Fabry-P\'erot with very low reflectivity) taking into account multiple reflections on its surfaces.
In \cite{ref:zhang} an accurate and efficient algorithm describing the electric field propagation in both isotropic and anisotropic etalons (and crystals in general) is presented. The study takes into consideration cross-talks between orthogonally polarized components and the effect of multi-layer coatings, but no analytical expressions are obtained. A general theory of anisotropic etalons describing its polarimetric properties has not been presented yet up to our knowledge. 

This is the second in our series on Fabry-P\'erot etalon-based instruments. After a comprehensive view of isotropic etalons (interferometers made with isotropic materials) and a discussion on the two most typical configurations for etalons in astronomical instruments, here we concentrate in the anisotropic case. We carry out an analytical and numerical study of the polarimetric properties of uniaxial crystalline etalons (also applicable to liquid crystal etalons), to evaluate the effect of the birefringence introduced when the ray direction and the optical axis of the crystals are not parallel. We will neglect the effect of multi-layer coatings for the sake of simplicity since their effect is expected to be many orders of magnitude smaller than that of the results presented here \citep{ref:doerr,ref:zhang}. 

First, we study the induced birefringence in crystalline etalons (Sect.~\ref{sec:birefringence}); second, we derive the Mueller matrix of the etalon (Sect.~\ref{sec:Mueller}); and then we focus on its polarimetric response (Sect.~\ref{sec:response}). Special emphasis is put into etalons in a telecentric configuration since misalignments appear in them in a natural way. We discuss the effects in the point spread function of the system (Sect.~\ref{sec:PSF}) and we analyze qualitatively the impact of birefringent etalons on solar instruments (Sect~\ref{sec:comments}). A thorough analysis on the consequences of using a Fabry-P\'erot on real instruments is considered on the next work of this series of papers. Finally, we draw the main conclusions (Sect.~\ref{sec:conclusions}).

\section{Birefringence induced in crystalline etalons} \label{sec:birefringence}

Crystalline uniaxial etalons present a given direction called {\em optical axis}, $\ee$, along which the two orthogonal components of the electric field stream with the same velocity. If the wavefront normal, $\s$, is parallel to $\ee$, then the orthogonal components of the electric field travel with the same velocity, as it happens for normal illumination in $Z$-cut crystalline etalons. In such a situation, birefringence effects are not present. However, in any other direction, the propagation of the electric field components should be studied separately because they travel across a medium with different refraction indices.

\begin{figure}[t]
\begin{center}
    \includegraphics[width=0.48\textwidth]{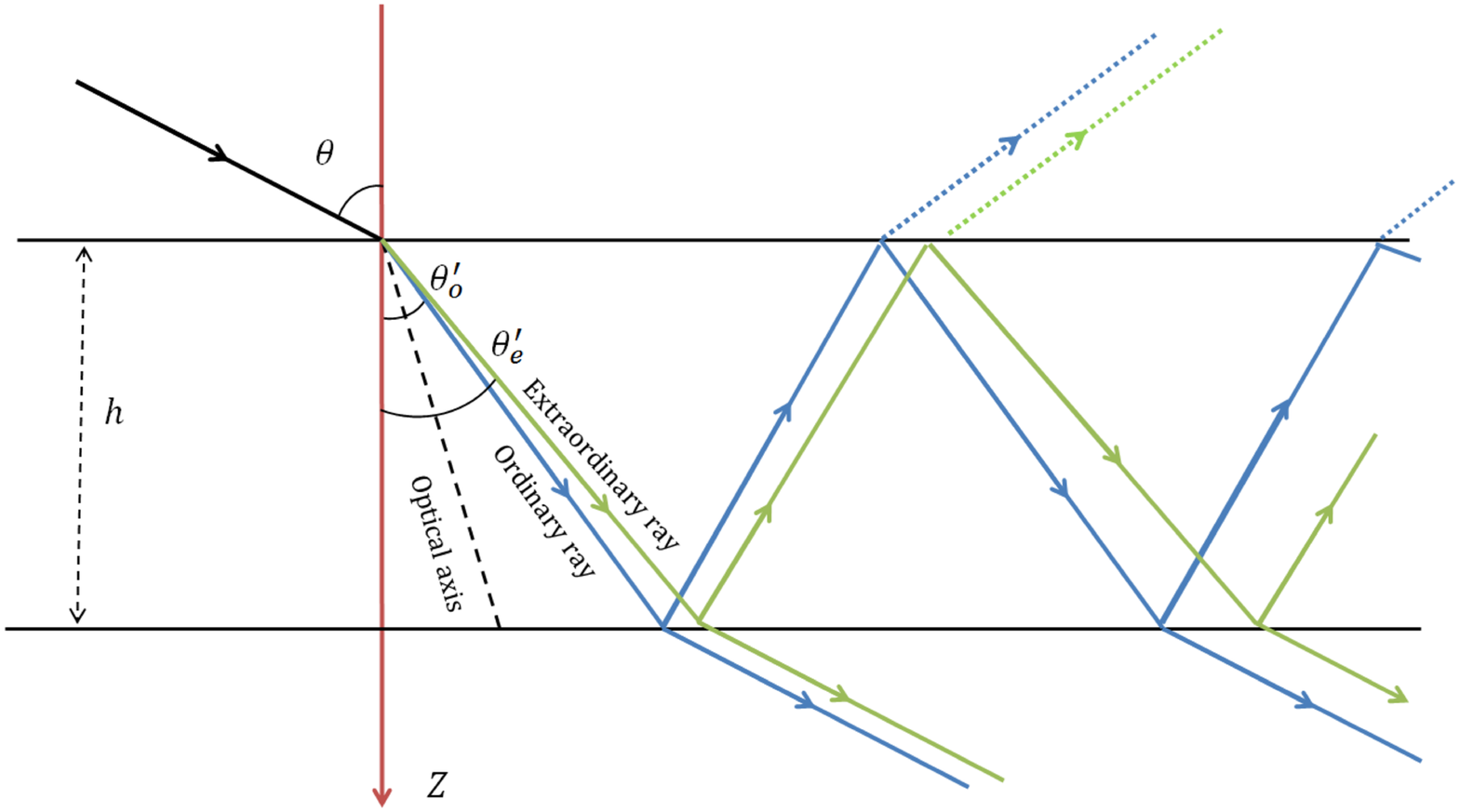}
	\caption{Layout of a ray with a certain incident angle $\theta$ entering an uniaxial etalon (black) whose optical axis is not parallel to the surface normal. (The convention for a $Z$-cut crystal calls $Z$ that normal, but we reserve $Z$ for the axis along the ray direction; see text for details.) The ray is split in two orthogonal rays, the ordinary (blue) and the extraordinary (green), each one refracted with different angles $\theta'_o\neq\theta'_e$ and thus having different optical paths and different phase when either transmitting or reflecting at the etalon surfaces.}
	\label{esquema}
\end{center}
\end{figure}

For any given ray direction, $\hat{\textbf t}$, the plane formed between $\ee$ and $\hat{\textbf t}$ is called the {\em principal plane} of the medium.\footnote{The principal plane is also defined as that containing the optical axis, $\ee$ and the wavefront normal, $\s$. Both definitions are equivalent since $\hat{\textbf t}$ and $\s$ are coplanar with $\ee$.} Then, the electric field vector can be considered as the sum of two incoherent orthogonally polarized components:

\begin{equation}
\textbf{E}=\textbf{E}_{o}+\textbf{E}_{e},
\end{equation}
where $\textbf{E}_{o}$ and $\textbf{E}_{e}$ are the so-called {\em ordinary} and {\em extraordinary} electric field components. The propagation of a light beam can be thought of as that of two linearly polarized beams, one having a velocity independent of direction, the ordinary beam, and the other with a velocity depending on direction, the extraordinary beam. The ordinary beam propagates like any beam through an isotropic medium. That is not the case for the extraordinary beam, whose energy does not propagate along the wavefront normal (but along the ray direction), unless this is parallel to the optical axis.

Figure \ref{esquema} shows the splitting of an incoming ray with incidence angle $\theta$ when traveling through an etalon with its optical axis misaligned with respect to the surface normal. The ordinary and extraordinary rays propagate along different directions and, thus, traverse different optical paths at the exit.

The first measurable effect of the different propagation of both rays is a phase difference between the ordinary and the extraordinary beams because they split, and behave independently, except if $\s\cdot\ee=1$. The difference in phase produced between every two successive extraordinary and ordinary beams due to its different geometrical paths through an etalon (Fig.~\ref{esquema}) is simply given by 
\begin{equation}
\varphi\equiv\delta_e-\delta_o=\frac{4\pi h}{\lambda}(n_e\cos\theta_e'-n_o\cos\theta_o'),
\label{varphi}
\end{equation}
where subindices $o$ and $e$ refer to the ordinary and extraordinary rays respectively.\footnote{We shall be using the basic nomenclature of Paper I for the sake of consistency. Hence, we refer the reader to that paper for the possible missing definitions.} 

Within an etalon, the wavefront direction vectors of the ordinary and extraordinary rays depend on the incident wavefront direction and on the refraction index for both the ordinary and extraordinary components.
The geometrical path along the ray and wavefront directions coincide for the ordinary ray but not for the extraordinary ray.
 Furthermore, the refraction index of the extraordinary beam depends on the direction $\s$, so the propagation of the extraordinary component is more complex than that of the ordinary beam \citep{ref:born}:

\begin{equation}
\frac{1}{n_e^2 (\gamma)}=\frac{1}{n_o^2}\cos^2\gamma+\frac{1}{n_3^2}\sin^2\gamma,
\label{nee}
\end{equation}
where $\gamma$ is the angle between $\s$ and $\ee$ and $n_3$ is the refraction index for an electric field vibrating along the optical axis of the etalon. Notice that $n_e=n_o$ if $\gamma=0$. 

The ordinary and extraordinary components propagate such that their transmitted electric field vectors can be given by Eq.~(45) of Paper I, each with their respective retardance:
\begin{equation}
\Et_o=\frac{\sqrt{\tau_o}}{1-R}\frac{{\rm e}^{{\rm i}\delta_o/2}-R{\rm e}^{-{\rm i}\delta_o/2}}{1+F\sin^2(\delta_o/2)}\Ei_o,
\label{Eo}
\end{equation}
\begin{equation}
\Et_e=\frac{\sqrt{\tau_e}}{1-R}\frac{\left({\rm e}^{{\rm i}[\delta_o+\varphi]/2}-R{\rm e}^{-{\rm i}[\delta_o+\varphi]/2}\right)}{1+F\sin^2\left([\delta_o+\varphi]/2\right)}\Ei_e,
\label{Ee}
\end{equation}
where $\tau_o$ and $\tau_e$ account for possible different values of the absorptance of the etalon for the ordinary and extraordinary rays. Note that even for $R=0$ a retardance $\varphi/2$ is induced between the ordinary and extraordinary rays.

Without loss of generality, to describe the electromagnetic field components, we can choose  a reference frame in which the $Z$ axis coincides with the ray direction, the Poynting vector direction (see Figure \ref{xyz}). This choice is kept for any incident ray. For the sake of simplicity, let us make the $X$ axis coincide with the direction of vibration of the ordinary electric field. The $Y$ axis is then parallel to the plane that contains the extraordinary electric field. In this reference frame, the transmitted electric field has only two orthogonal components, $\Et_x$ and $\Et_y$, whose propagation can be expressed in matrix form as

\begin{equation}
\begin{pmatrix}
\Et_x\\
\Et_y
\end{pmatrix}
=\Hmatrix
\begin{pmatrix}
\Ei_o\\
\Ei_e
\end{pmatrix},
\label{Jones}
\end{equation}
where $\Hmatrix$ is the so-called Jones matrix. Since the ordinary and extraordinary components are orthogonal and behave independently, $\Hmatrix$ is diagonal in the chosen reference frame. Their components are given by the factor relating $\Ei$ and $\Et$ in  Equations (\ref{Eo}) and (\ref{Ee}). Usually, an arbitrary choice of the $X$ and $Y$ directions will not coincide  with the plane containing $\Ei_o$ and $\Ei_e$, because the principal plane orientation depends on both the optical axis and ray directions. In that case, a rotation of the reference frame about $Z$ is needed, as discussed in-depth in Section \ref{sec:rotations}.

Equation \ref{Jones} is valid only for the propagation of $\Ei_o$ and $\Ei_e$ in a beam strictly collimated. As pointed out in Paper I, to obtain an expression valid for converging illumination, we have to integrate both the ordinary and extraordinary rays all over the aperture of the beam (the pupil in case of telecentric illumination) and get $\Etilde_o$ and $\Etilde_e$ (see Eq. [48] of Paper I). Then, it is easily seen that an equation like

\begin{equation}
\begin{pmatrix}
\Etilde_x\\
\Etilde_y
\end{pmatrix}
=\Htilde\prima
\begin{pmatrix}
\Ei_o\\
\Ei_e
\end{pmatrix}
\label{Jones2}
\end{equation}
can be written, where the linearity of the problem yields the new Jones matrix elements, $\Htildeel\prima_{ij}$, as direct integrals of the old ones. Since the principal plane differs for each particular ray direction, a rotation of the Jones matrix needs also to be added, as thoroughly explained in Section \ref{tele}.

\section{Mueller matrix for crystalline etalons} \label{sec:Mueller}
\subsection{General expression}
Due to the birefringence induced by anisotropic crystalline etalons, uniaxial Fabry-P\'erot filters show different responses for each of the incoming Stokes vector components. The more general way to study the polarization response of these etalons is by using the Mueller matrix formulation. According to \cite{ref:jefferies}, the elements of the Mueller matrix, ${\rm M_{ij}}$, are given by

\begin{equation}
{\rm M_{ij}}=\frac{1}{2}{\rm Tr}\left[\sigma_i{\Hmatrix}\sigma_{j}{\Hmatrix}^\dagger\right],
\label{eq:jefferies}
\end{equation}
where $\sigma_i$ ($i=0,1,2,3$) are the identity and Pauli matrices with the sorting convention employed in \cite{ref:spectropolarimetry}. In case $\Et_x$ and $\Et_y$ are parallel to $\Ei_o$ and $\Ei_e$, as described in the previous section, ${\Hmatrix}$ is diagonal and the Mueller matrix can be expressed in the form
\begin{equation}
\Mmatrix=
\begin{pmatrix}
a & b & 0 & 0\\
b & a & 0 & 0\\
0 & 0 & c & -d\\
0 & 0 & d& c
\end{pmatrix},
\label{M}
\end{equation}
whose coefficients are given by
\begin{equation}
\begin{gathered}
a=\frac{1}{2}(\H_{11}\H_{11}^\ast+\H_{22}\H_{22}^\ast),\\
b=\frac{1}{2}(\H_{11}\H_{11}^\ast-\H_{22}\H_{22}^\ast),\\
c=\frac{1}{2}(\H_{22}\H_{11}^\ast+\H_{11}\H_{22}^\ast),\\
d=\frac{\rm i}{2}(\H_{22}\H_{11}^\ast-\H_{11}\H_{22}^\ast),
\end{gathered}
\label{abcd}
\end{equation}
where $*$ refers to the complex conjugate. Using basic trigonometric equivalences
and defining
\begin{equation}
 \taueff\equiv\sqrt{\tau_o\tau_e},
 \end{equation}
 \begin{equation}
 \bar{\tau}\equiv\frac{\tau_o+\tau_e}{2},
 \end{equation}
  it can be deduced (Appendix \ref{Aa}) that

\begin{align}
&a=\frac{\taueff}{\zeta}\left[\frac{\bar{\tau}}{\taueff}+\Gamma\right],\\
&b=\frac{\taueff}{\zeta}\Lambda,\\
&c=\frac{\taueff}{\zeta}\left[\Psi+\frac{\cos(\varphi/2)}{(1-R)^2}\right],\\
&d=\frac{\taueff}{\zeta}\left[\Omega-\frac{\sin(\varphi/2)}{(1-R)^2}\right],
\label{abcd2}
\end{align}

where
\begin{align}
\zeta=&\left[1+F\sin^2\left(\frac{\delta_o}{2}\right)\right]\left[1+F\sin^2\left(\frac{\delta_o+\varphi}{2}\right)\right],
\label{zeta}\\
\Gamma=&\frac{F}{2\taueff}\left[\tau_o\sin^2\left(\frac{\delta_o+\varphi}{2}\right)+\tau_e\sin^2\left(\frac{\delta_o}{2}\right)\right],
\label{alpha_coeff}\\
\Lambda=&\frac{F}{2\taueff}\left[\frac{\tau_o-\tau_e}{F}+\tau_o\sin^2\left(\frac{\delta_o+\varphi}{2}\right)-\tau_e\sin^2\left(\frac{\delta_o}{2}\right)\right],
\label{beta}\\
\Psi=&\frac{F}{4}\left[R\cos\left(\frac{\varphi}{2}\right)-2\cos\left(\delta_o+\frac{\varphi}{2}\right)\right],
\label{gamma}\\
\Omega=&-\frac{F}{4}R\sin\left(\frac{\varphi}{2}\right).
\label{sigma}
\end{align}
Notably,
\begin{equation}
a+b=\frac{\tau_o}{1+F\sin^2\left(\delta_o/2\right)},
\label{aplusb}
\end{equation}
and
\begin{equation}
a-b=\frac{\tau_e}{1+F\sin^2\left(\delta_e/2\right)}.
\label{aminusb}
\end{equation}
That is, the transmission profiles (Eq. 11 in Paper I) for the ordinary and extraordinary rays are recovered from the sum and subtraction of the two first elements of the Mueller matrix.

The Mueller matrix of a birefringent etalon is expressed as a function of the etalon parameters and the retardance induced between the ordinary and extraordinary rays. We can separate it into two matrices, one similar to that describing an ideal retarder, $\Mmatrix_{r}$, and another one as a mirror due to the fringing effects, $\Mmatrix_{m}$:
\begin{equation}
\Mmatrix=\Mmatrix_{r}+\Mmatrix_{m}.
\label{M2}
\end{equation}
If we define $\tauprima\equiv\taueff(1-R)^{-2}$, we find that
\begin{equation}
\Mmatrix_{r}=\frac{\tauprima}{\zeta}
\scalebox{.9}{$
\begin{pmatrix}
\bar{\tau}\taueff^{\prime -1} & 0 & 0 & 0 \\
0 & \bar{\tau}\taueff^{\prime -1} & 0 & 0\\
0 & 0 & \cos(\varphi/2) & \sin(\varphi/2)\\
0 & 0 & -\sin(\varphi/2) & \cos(\varphi/2)
\end{pmatrix}$},
\label{Mr}
\end{equation}
and

\begin{equation}
\Mmatrix_{m}=\frac{\taueff}{\zeta}
\scalebox{.9}{$
	\begin{pmatrix}
	\Gamma & \Lambda & 0 & 0\\
	\Lambda & \Gamma & 0 & 0\\
	0 & 0 & \Psi & -\Omega\\
	0 & 0 & \Omega & \Psi
	\end{pmatrix}$}.
\label{R}
\end{equation}
The extraordinary direction in our numerical examples coincides with the fast axis, $Y$, since we set $n_e<n_o$.

It is also worth noticing that both matrices are multiplied by $\zeta^{-1}$, which depends on both $\delta_0$ and $\varphi$. Since these two quantities are wavelength and direction dependent, Eq.~(\ref{M2}) does not strictly correspond to the sum of a retarder and a mirror, except in the collimated, monochromatic case. In the limit when $R=0$, since $F\equiv4R(1-R)^{-2}$ (Eq. 13 of Paper I), the mirror matrix vanishes and the etalon Mueller matrix turns into that of an ideal retarder. On the other hand, in the limit when $\varphi=0$ , i.e., in the limit of an isotropic etalon, it can be shown that $\Mmatrix$ is reduced to the identity matrix except for a proportionality factor that corresponds to the transmission factor of an isotropic etalon (Eq. 11 in Paper I). 

\cite{ref:lites} obtained a similar expression than Eq.~(\ref{M2}) but restricted to  $R<\!\!<1$ and assuming both normal incidence on the etalon and that the optical axis is perpendicular to the surface normal. In that case the dependence on the wavelength and direction disappears and matrices in Eq.~(\ref{M2}) describe an ideal retarder and an ideal mirror. Our result is completely general since it is valid for any value of $R$ and for any incident angle of the wavefront. Also notice that in \cite{ref:lites} the plus and minus signs of $d$ are interchanged due to the sign convention in the definition of the harmonic plane waves, and, therefore, in $\sigma_3$.

Whenever $\varphi\neq0$, the Mueller matrix becomes non-diagonal and spurious signals in the measured Stokes vector, known as {\em polarization cross-talk} in the solar physics jargon, are introduced. This is so because  $b$  and $d$ introduce in Eq.~(\ref{M}) cross-talk signals between $I$ and $Q$ and between $U$ and $V$ respectively if the Stokes parameters are measured after passing the light through the etalon. There are, however, some cases where these cross-talks are not relevant. For example, for totally polarized light in the $Q$ direction, $I^{(i)}=Q^{(i)}$ and $U^{(i)}=V^{(i)}=0$ and, therefore
\begin{equation}
I^{(t)}=Q^{(t)}=(a+b)I^{(i)}=(a+b)Q^{(i)}.
\label{eq:IeqQ}
\end{equation}
Hence, the transmission equation of an (ideal) isotropic etalon with the ordinary refractive index is recovered. This happens, for example, if the etalon is located after a linear polarizer with its optical axis parallel to the $+Q$ direction. In this case, artificial polarization signals do not appear. 

So far, we have restricted to collimated illumination of the etalon with a convenient reference frame for expressing the Stokes vector. In telecentric configuration, the shape of the Mueller matrix, $\tilde{\Mmatrix}$, is the same in practice. The matrix elements, however, have a much more involved expression than in Eqs.~(\ref{abcd}) and (\ref{M2}) because they are now calculated from the integrated Jones matrix, $\Htilde$, across the pupil. This will be discussed in detail in Section \ref{tele}.

\subsection{Rotations of the Mueller matrix}
\label{sec:rotations}
\begin{figure}[t]
	\includegraphics[width=0.52\textwidth]{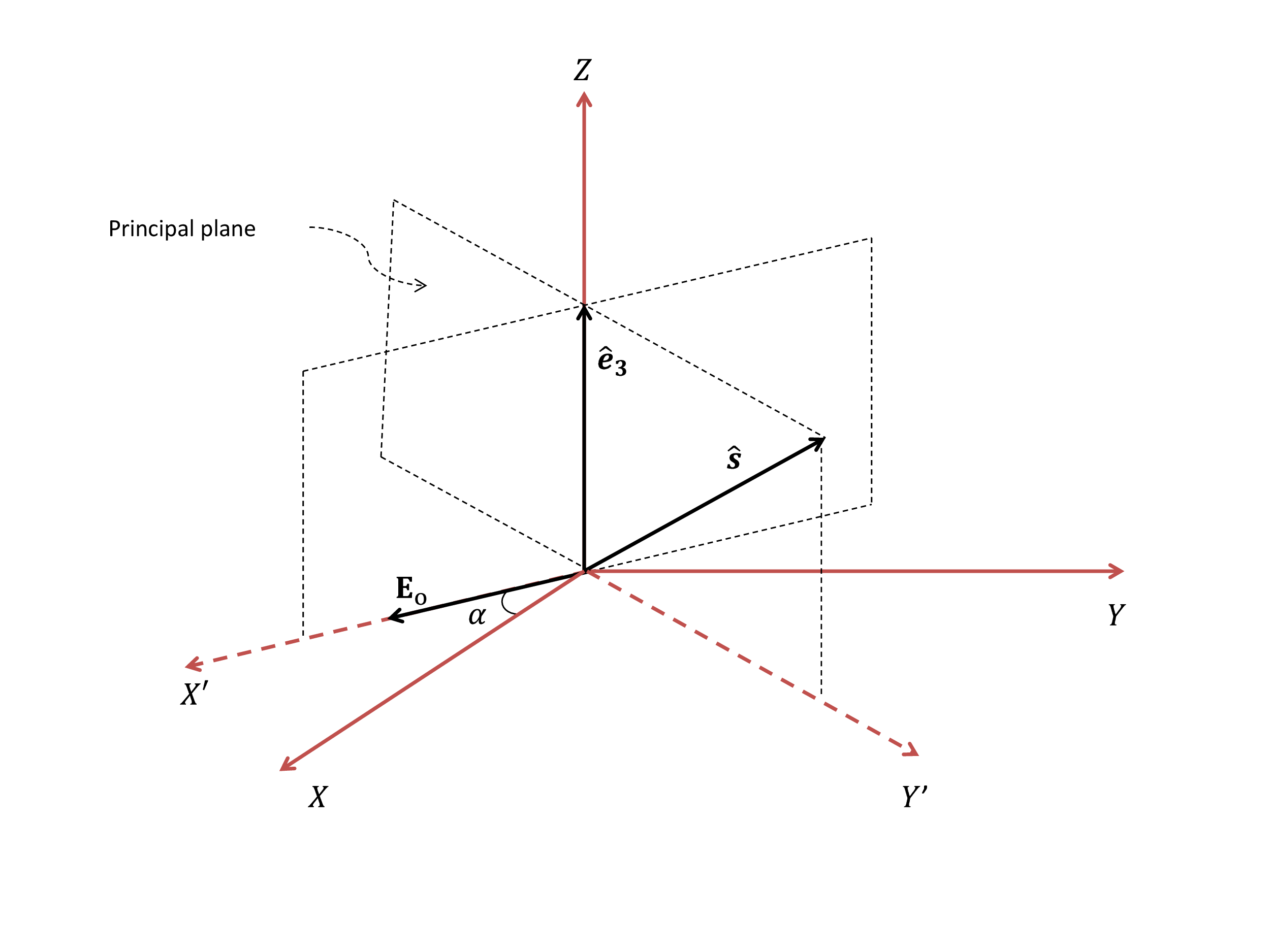}
	\caption{General reference frame $XYZ$ where the ray direction $\hat{\textbf t}$ of the etalon coincides with $Z$. The ordinary electric field component, $\rm \textbf{E}_o$, is contained on a plane perpendicular to the principal plane and forms an angle $\alpha$ with $X$ which depends on the $\hat{\textbf e}_3$ direction. Spherical coordinates to describe the wavefront direction unitary vector $\hat{\textbf s}$ are also included: $\beta$ is the polar angle (measured from Z) and $\phi$ is the azimuthal angle (measured from $X$).}
	\label{xyz}
\end{figure}

Since we here restrict to $Z$-cut crystals, propagation through the normal to the etalon reflecting surfaces does not produce any birefringence effect if the optical axis is perfectly aligned. For normal illumination, the choice of the $\mathbf{+Q}$ direction is, then, irrelevant, since both the ordinary and extraordinary rays travel with the same velocity. For any other incident angle, a careful choice of the $+Q$ direction must be made, though. From Eq.~(\ref{Jones}) it is natural to take the ordinary electric field direction ($+Q$ direction) as the $X$ axis. The geometry of the problem is depicted in Fig.~\ref{xyz}, where $Z$ represents the ray direction. This direction does not necessarily coincide with the optical axis, $\hat{\textbf{e}}_3$.Then, $\E_o$ vibrates in a plane perpendicular to the principal plane \citep[see, for example,][]{ref:spectropolarimetry}. A rotation of an angle $\alpha$ about $Z$ is then mandatory for the Mueller matrix of the etalon to give proper account of birefringence. No further rotations are needed, though, since our reference frame is chosen such that $Z$ coincides with the direction of observation. 

The rotation angle $\alpha$ is given by
\begin{equation}
\cos\alpha=\textbf{u}_x \cdot \textbf{u}_{x'},
\label{angle}
\end{equation}
where $\textbf{u}_x=(1,0,0)$ is the unitary direction vector of $X$ and $\textbf{u}_{x'}$ is the unitary direction vector of $X'$, which may be calculated  from the normalized vectorial product of $\hat{\textbf t}=(0,0,1)$ and $\hat{\rm \textbf{e}}_3=\left(e_3^{(x)},e_3^{(y)},e_3^{(z)}\right)$:
\begin{equation}
\textbf{u}_{x'}=\left(e_3^{(y)},-e_3^{(x)},0\right).
\end{equation}
If we use polar, $\beta$, and azimuthal, $\phi$, angles to describe $\hat{\textbf e}_3$ (Fig.~\ref{xyz}), it is easy fo find that
\begin{equation}
\cos\alpha=\sin\phi.
\end{equation}
 Thus,
\begin{equation}
\alpha=\pm(\phi-\pi/2).
\label{alpha}
\end{equation}
This equation is valid whenever the angle between the ray direction and the optical axis, $\beta$, is different from zero. If $\beta=0$, the wavefront normal and the optical axis are parallel and we can set arbitrarily $\alpha=0$ because of the rotational symmetry of the etalon about $Z$. It is important to remark that the polar and azimuthal angles are uncoupled in our description. That is, the retardance $\varphi$ between the ordinary and extraordinary rays only depends on the angle $\beta$; whereas the rotation angle, $\alpha$, only depends on $\phi$. 

The dual solution for $\alpha$ in Eq.~(\ref{alpha}) reflects the intrinsic $180\degree$ ambiguity in polarimetry as the situation described so far would be exactly the same for an ordinary electric field $-\E_o$. The usual convention is to employ positive signs for counterclockwise rotations (right-handed) and the negative sign for clockwise rotations. Consequently, if we set the $XYZ$ directions as our reference frame, we should rotate the Mueller matrix of the etalon an angle $\alpha=\phi-\pi/2$ and vice versa. The Mueller matrix can then be cast as
\begin{equation}
\Mmatrix_{\alpha}=
\scalebox{1}{$
\begin{pmatrix}
a & bC_2 & bS_2 & 0\\
bC_2 & aC_2^2+cS_2^2 & (a-c)S_2C_2 & dS_2\\
bS_2 & (a-c)S_2C_2 & aS_2^2+cC_2^2 & -dC_2\\
0 & -dS_2 &dC_2 & c\\
\end{pmatrix}$},
\label{Mrot}
\end{equation}
where the coefficients $C_2$ and $S_2$ are given by $C_2=\cos2\alpha$, $S_2=\sin2\alpha$. This matrix gathers all the necessary information to describe the propagation of the Stokes components of any incident ray in the etalon.

\section{Polarimetric response of birefringent etalons} \label{sec:response}

In any linear system, the polarimetric response is determined by its Mueller matrix coefficients, which are independent from the incident Stokes vector. We have shown that the Mueller matrix of crystalline etalons \emph{only} depends on four independent coefficients, and on the azimuthal orientation of the principal plane. The coefficients are related to optical parameters of the etalon (e.g., refraction indices and geometrical thickness), to the wavelength and to the phase difference between the extraordinary and ordinary beams. The phase difference depends, at the same time, on the relative direction of the ray direction with respect to the optical axis. In this section we study the spectral behavior of these parameters in three different cases: (1) when collimated light illuminates the etalon with a certain angle with respect to the normal; (2) when the illumination is normal to the etalon but the optical axis is misaligned; and (3), when the etalon is illuminated in telecentric configuration. We use the parameters of SO/PHI $Z$-cut LiNbO$_3$ etalon: $n_o=2.29$, $n_3=2.20$, $h=251.63$ $\rm\mu m$, $A=0$, $R=0.92$ and $\lambda_0=617.3356$ nm. We also assume the etalon is immersed in air and that $\tau_o=\tau_e=1$ for simplicity. The results can easily be extended to any etalon based on uniaxial crystals.

In LiNbO$_3$, the birefringence is typically smaller than $0.1$ \citep[e.g.,][]{ref:nikogosyan} and can be neglected compared to $n_o$ and $n_e$. Hence, a compact analytical expression for $\varphi$ as a function of the incident angle and of the angle formed by the optical axis with $Z$ can be found. Specifically, it can be shown \citep{ref:born} that
\begin{equation}
n_e\cos\theta_e'-n_o\cos\theta_o'\simeq\frac{1}{\cos\theta'}(n_3-n_o)\sin^2(\theta'-\theta_3),
\end{equation} 
where $\theta_3$ is the angle between the optical axis and the surface normal, and $\theta'$ is an arbitrary fictitious refracted angle that is given by

\begin{figure*}
\centering
\includegraphics[width=0.9\textwidth]{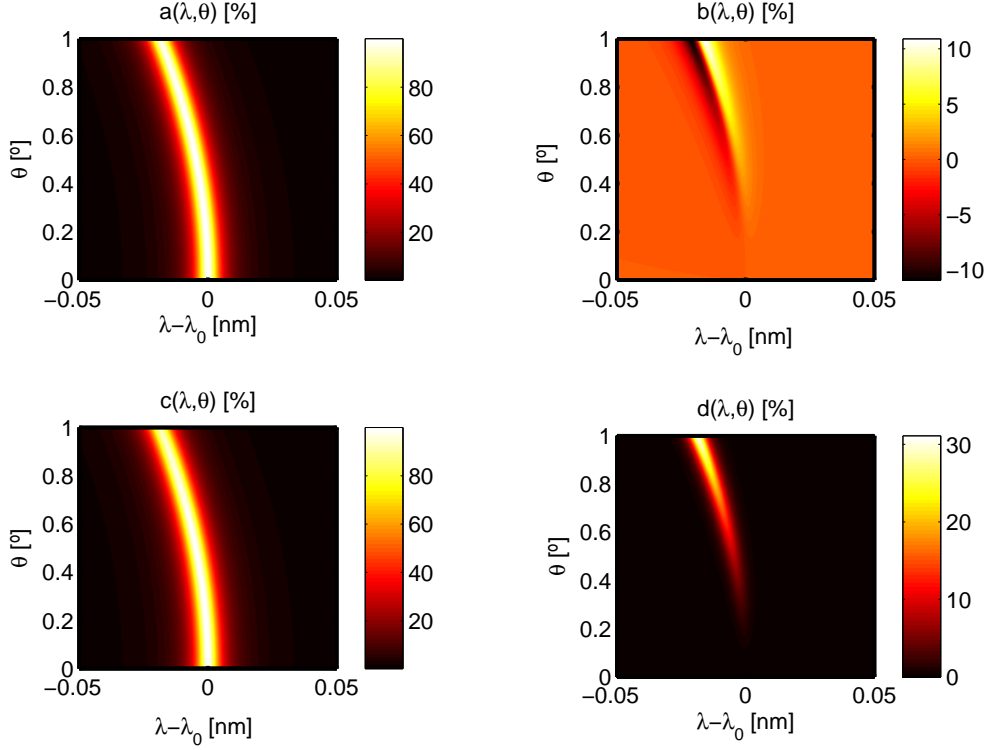}
	\caption{Variation of the $a,b,c$ and $d$ coefficients of the Mueller matrix of the etalon as a function of wavelength and incident angle.}
	\label{abcd2d}
\end{figure*}

\begin{equation}
\theta'=\sin^{-1}\left(\frac{n\sin\theta}{n\prima}\right),
\end{equation}
where $n$ is the refraction index of the medium in which the etalon is immersed and $n\prima$ can be taken as  the average between the $n_3$ and $n_o$: 
\begin{equation}
n\prima\equiv\frac{n_3+n_o}{2}.
\label{na}
\end{equation}
 Thus, we can rewrite Eq.~(\ref{varphi}) as
\begin{equation}
\varphi\simeq\frac{4\pi h}{\lambda\cos\theta'}(n_3-n_o)\sin^2(\theta'-\theta_3),
\label{varphi1}
\end{equation}
which directly depends on $\theta_3$ and $\theta'$. In the case $\theta'$ and $\theta_3$ are close to zero, we can further approximate this expression as
\begin{equation}
\varphi\simeq\frac{8\pi h}{\lambda(2\nprimac-n^2\theta^2)}(n_3-n_o)(n\theta-n\prima\theta_3)^2,
\label{varphi_aprox}
\end{equation}
which is expressed as a function of the incident angle instead of the fictitious refracted angle. It is important to notice that when $\theta$ and $\theta_3$ are zero, $\varphi=0$ as predicted for normal illumination. Whenever either $\theta$ or $\theta_3$ are different from zero, birefringent effects appear on the etalon. On the other hand, retardance increases with the width of the etalon, with the birefringence of the crystal and the inverse of the wavelength (it is therefore larger in the ultraviolet than in the infrared region). It is important to remark that Eq.~(\ref{varphi1}) is an approximate expression valid for materials with small birefringence. An exact formula of the retardance without restrictions in the magnitude of the birefringence was found by \cite{ref:veiras}. The validity of Eq.~(\ref{varphi1}) in our numerical examples is discussed in Appendix \ref{Exact_phase}.

\subsection{Effect of oblique illumination in collimated etalons} \label{Oblique}
Consider a perfectly parallel and flat etalon with its optical axis aligned with the normal to the reflecting surfaces. Let us illuminate it with a collimated monochromatic beam with incidence angle $\theta$. We assume we have chosen a reference frame in which $\alpha = 0$, so the etalon Mueller matrix is given by Equation (\ref{M}). Then, the etalon will behave as a wavelength-dependent retarder plus a mirror as described in Eq.~(\ref{M2}), modifying the polarization properties of the incoming Stokes vector. To see the effects, we represent the variation of the Mueller matrix coefficients as a function of the incident angle in Figure \ref{abcd2d}. We have restricted $\theta$ to vary from $0^{\circ}$ to 1$\degree$ and we have limited the spectral range to the region $\lambda_0\pm\delta\lambda$, where $\delta\lambda=0.05$  ${\rm nm}$, to cover the whole transmission profile centered at $\lambda_0$ (the location of the maximum transmission for normal incident illumination). 

\begin{figure}[t]
\includegraphics[width=0.52\textwidth]{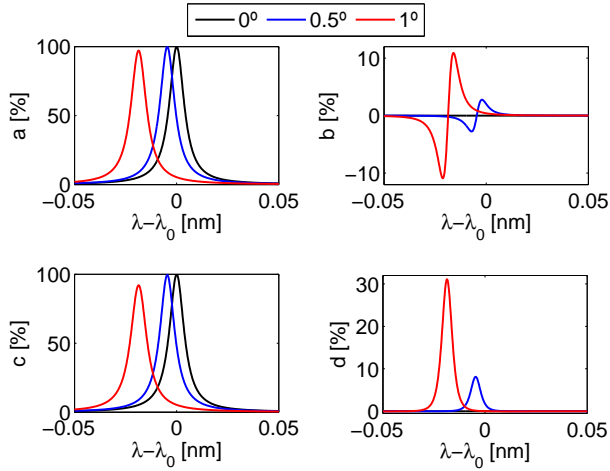}
\caption{$a,b,c$ and $d$ coefficients of the Mueller matrix of the etalon as a function of wavelength for incident angles $0\degree$ (black solid line), $0\fdeg5$ (blue solid line) and $1\degree$ (red solid line).}
\label{abcd1d}
\end{figure}

\begin{figure}
\includegraphics[width=0.45\textwidth]{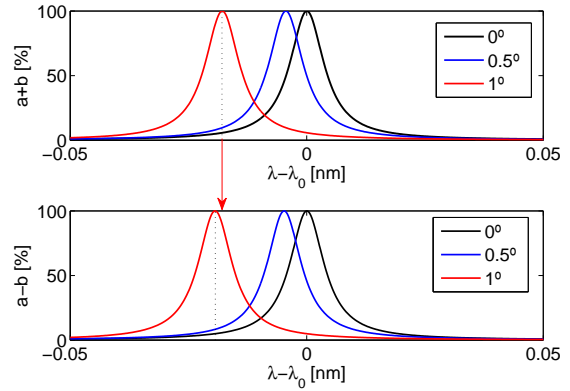}
\caption{Spectral dependence of the transmission profile of the ordinary ray ($a+b$) and of the extraordinary ray ($a-b$) for different incidence angles ($\theta_3=0$): $0\degree$ (black solid line), $0\fdeg5$ (blue solid line) and $1\degree$ (red solid line). The vertical solid lines pinpoint the peak location for the $1\degree$ case in both panels. Notice that the peaks are located at different wavelengths, as explained in the text.}
\label{a+b}
\end{figure}

\begin{figure*}
	\plottwo{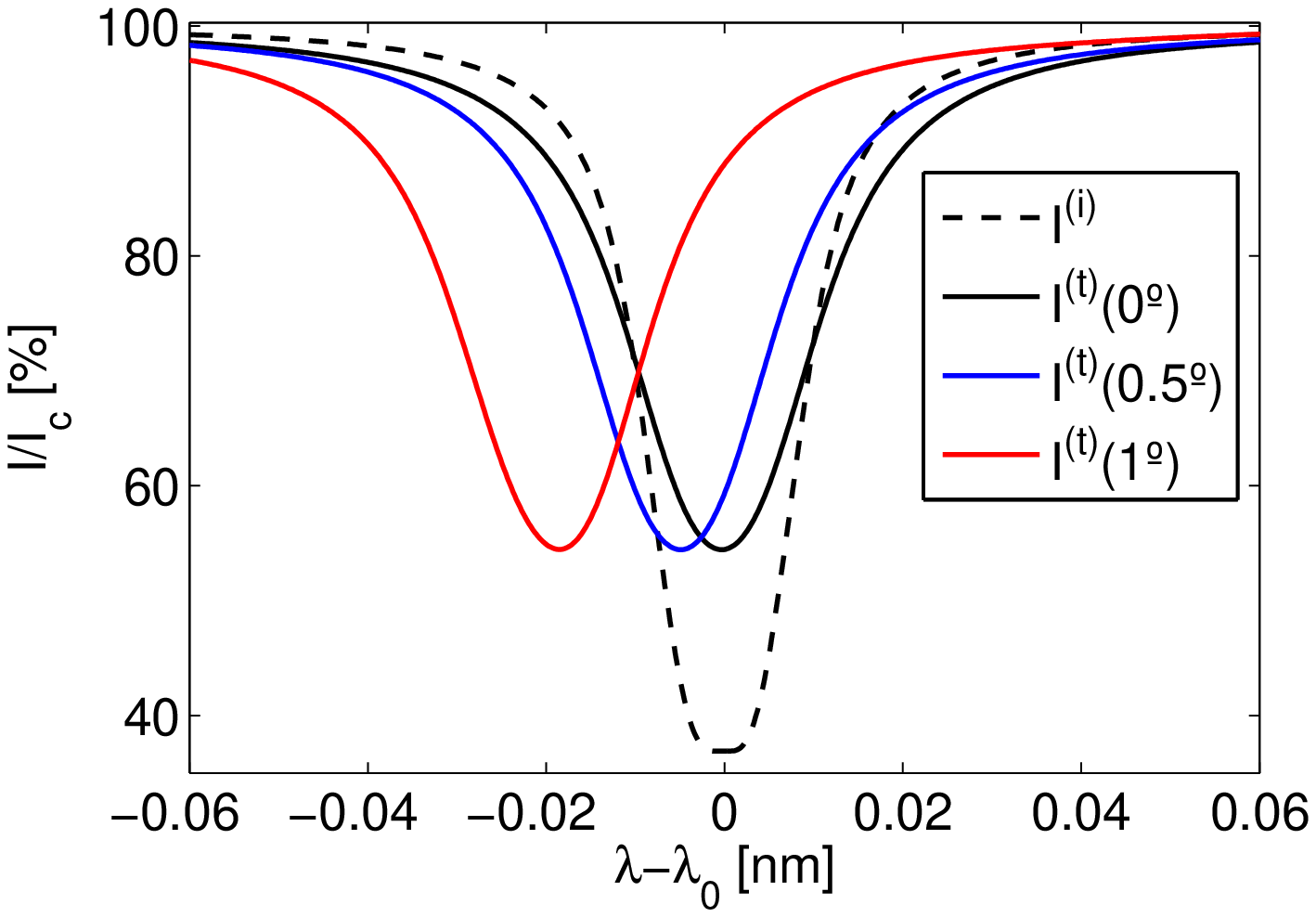}{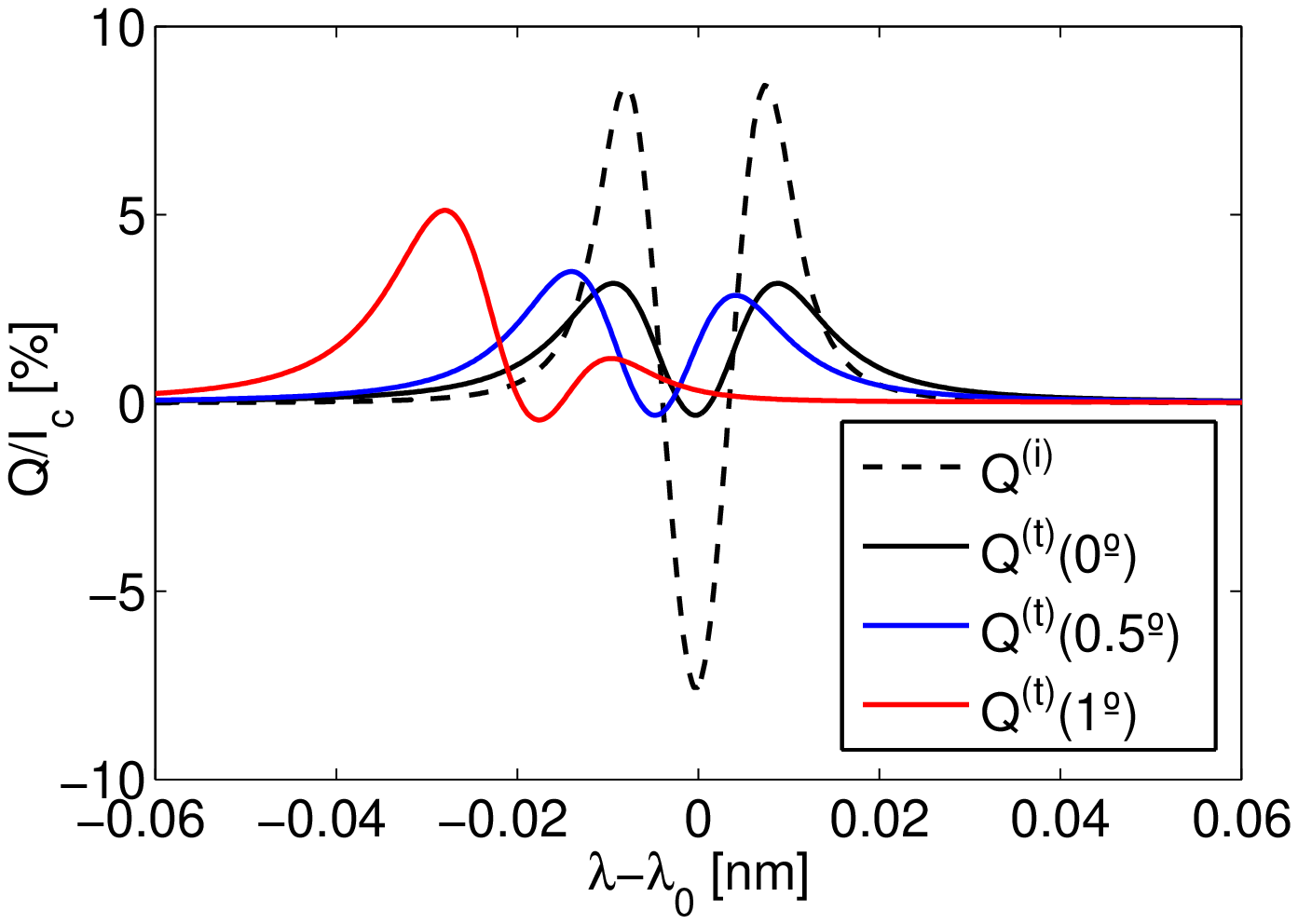}
	\caption{Reconstructed intensity profile of a synthetic Stokes profile (black dashed line) with the spectral response of the etalon at $\theta=0$ (black solid line), $\theta=0\fdeg5$ (blue solid line) and $\theta=1\degree$ (red solid line).}
	\label{convI}
\end{figure*}
Notice that, at normal incidence, coefficients $a$ and $c$ are strictly the same and represent the \emph{monochromatic} transmission profile of a perfect etalon while $b$ and $d$ are just zero. That is, no cross-talks between Stokes parameters appear. As soon as $b$ and $d$ differ from zero, hence as soon as the incidence angle is larger than zero, cross-talks from $I$ to $Q$, from $Q$ to $I$, from $U$ to $V$, and from $V$ to $U$ appear. In typical solar observations, the second and third contaminations are less important because the orders of magnitude of the Stokes profile signals usually are ${\cal O}(I)>{\cal O}(Q,U,V)$. To get a better insight on the relative effects of birefringence, we have plotted cuts of the images in Fig. \ref{abcd1d} at incidence angles of $0\degree$, $0\fdeg5$ and $1\degree$. Already apparent in Fig. \ref{abcd2d}, there is a clear, non-linear wavelength shift of the four parameters with increasing $\theta$, as well as a decrease in the peaks of $a$ and $c$. This is due to the wavelength splitting of the ordinary and extraordinary rays that can hardly be seen in these plots but will become apparent in the next Section. 

Parameters $b$ and $d$ are different from zero only when $\theta\neq 0$, as expected. Since they correspond to the off-diagonal elements of the etalon Mueller matrix, they introduce cross-talk signals in the transmitted Stokes vector. Remarkably, these spurious signals may be as much as 10$\%$ in Stokes $Q$ (crosstalk from Stokes $I$ to Stokes $Q$ and vice-versa, see Fig.~\ref{abcd1d}) and up to 30$\%$ between Stokes $U$ and Stokes $V$. All coefficients are positive, except for $b$, whose sign and magnitude depend on the separation of the ordinary and extraordinary peak wavelengths (Equation~\ref{abcd}). The exact antisymmetric shape with respect to the peak wavelength of the $b$ coefficient heralds a wavelength splitting between the ordinary and extraordinary transmission profiles. Remember that these profiles coincide with $a+b$ and $a-b$, respectively, according to Eqs.~(\ref{aplusb}) and (\ref{aminusb}). Both the sum and the subtraction of these coefficients have also been plotted in order to check this property in Fig.~\ref{a+b}, where we can see that $a+b$ and $a-b$ profiles are symmetric and that both peak at different wavelengths. 


So far, we have examined a flat wavelength spectrum (i.e., a continuum) of the incident light beam. However, when variations of the intensity with wavelength exist, as naturally occurs in solar absorption lines, an explicit dependence on the etalon Mueller matrix with wavelength appears. As a consequence, the cross-talk introduced between the Stokes parameters is wavelength dependent. Indeed, as we will see, the etalon can introduce asymmetries in the observed Stokes profiles, even when the input Stokes profiles are symmetric with respect to the central wavelength of the line. Figure~\ref{convI} shows an example of what happens when we illuminate the etalon at different angles with synthetic Stokes $I$ and $Q$ profiles corresponding to the \ion{Fe}{1}~line at 617.3~nm. Again, we assume three different angles of incidence, $\theta=0\degree$, $\theta=0\fdeg5$ and $\theta=1\degree$. The observed Stokes profiles have been determined by using the expressions
\begin{equation}
I^{(t)}=\frac{1}{N}(a*\Iin+b*\Qin),
\label{Iconv} 
\end{equation}
\begin{equation}
Q^{(t)}=\frac{1}{N}(a*\Qin+b*\Iin),
\label{Qconv}
\end{equation}
where $*$ is the convolution operator and $N$ is a constant introduced to normalize the observed profile to the continuum given by
\begin{equation}
N=\int_{0}^{\infty}\!\!\!\!\!\!a(\lambda)\,{\rm d}\lambda.
\end{equation}

As expected, the observed Stokes $I$ and $Q$ profiles are broader and shallower in the case of Stokes $I$ and weaker in case of Stokes $Q$ than the synthetic ones due to the limited spectral bandwidth of the etalon. Moreover, they are both blue shifted with respect to the $\lambda_0$, as expected (see Paper I). Remarkably, the cross-talk from Stokes $I$ to Stokes $Q$, governed by the $b$ coefficient of the etalon Mueller matrix, introduces a clear asymmetry in the observed Stokes $Q$ profile. The asymmetries are evident when the incident angle $\theta$ is $1\degree$. These asymmetries are also present in Stokes $I$ because of the cross-talk from Stokes $Q$ to Stokes $I$, although in less amount due to the larger values of the incident Stokes $I$ component. In this case we have taken into account only linear polarized light, but the effects are larger when there is cross-talk between Stokes $U$ and Stokes $V$, as one can deduce by just looking at Figure~\ref{abcd1d}.

\subsection{Effect of local domains in the etalon}
\label{sec:domains}
Either during the manufacturing process of etalons or when applying an intense electric field, local domains where the optical axis is not perpendicular to the etalon surfaces may appear. This implies that, even when illuminating the etalon with a collimated beam normal to the etalon surfaces, birefringent effects can arise. The magnitude of these depend on the angles $\theta$ and $\theta_3$, as well as on the relative orientation $\alpha$ of the plane containing the ordinary ray electric field with respect to the chosen $X$ direction.

\begin{figure*}
\centering
\includegraphics[width=0.9\textwidth]{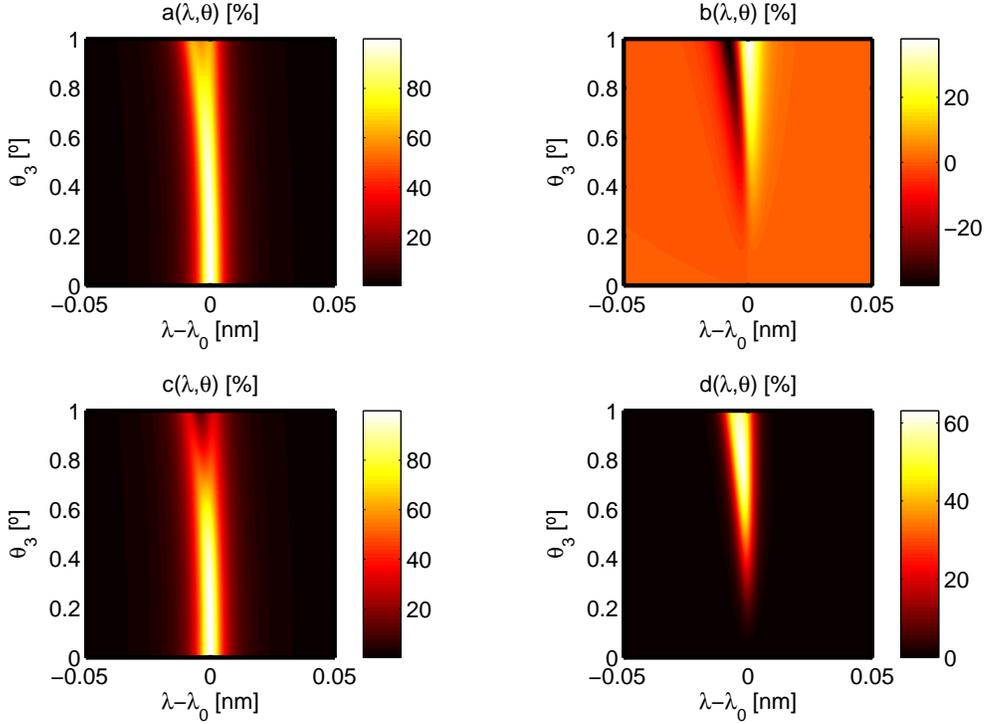}
	\caption{Variation of the $a,b,c$ and $d$ coefficients of the Mueller matrix of the etalon as a function of the wavelength and of $\theta_3$.}
	\label{theta3_2d}
\end{figure*}

\begin{figure}
\includegraphics[width=0.52\textwidth]{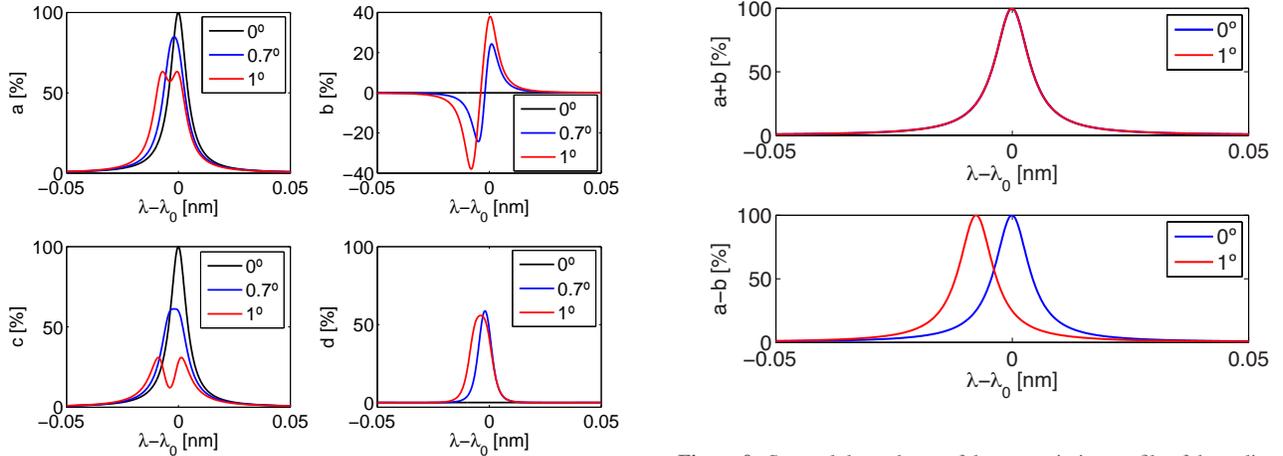}
\caption{Variation of the $a,b,c$ and $d$ coefficients of the Mueller matrix of the etalon as a function of the wavelength for $\theta_3=0^\circ$ (black solid line), $0.7^\circ$ (blue solid line) and $1^\circ$ (red solid line).}
	\label{theta3_1d}
\end{figure}

Let us suppose that we illuminate with a normal incident beam the etalon and that no other polarizing elements are present in our system. We can freely choose the $X$ axis of the etalon again to coincide with the vibration plane of $\E_o$. In this case, the Mueller matrix of the system is given again by Eq.~(\ref{M}) and 

\begin{equation}
\varphi=\frac{4\pi h}{\lambda}(n_3-n_o)\sin^2\theta_3,
\end{equation}
which, for small deviations, takes the form
\begin{equation}
\varphi\simeq\frac{4\pi h}{\lambda}(n_3-n_o)\theta_3^2.
\end{equation}
The non-zero elements of the Mueller matrix are plotted as a function of $\theta_3$ in Figures \ref{theta3_2d} and \ref{theta3_1d}. We can observe a similar effect than the one that appears when illuminating a perfect etalon with a collimated and oblique beam, except for the fact that, in this case, the spectral profile of the ordinary ray does not deviate and the splitting of the ordinary and extraordinary rays is more prominent. The spectral separation of the ordinary and extraordinary rays is clearly noticed in Fig.~\ref{theta3_2d} when approaching to $1\degree$. The double peak in the spectral transmission is also visible in $a$ and $c$ in Figure \ref{theta3_1d}. Cross-talks parameters $b$ and $d$ have a maximum absolute value of about $40\%$ and $60\%$, respectively.  

\begin{figure}[t]
	\begin{center}
		\includegraphics[width=0.45\textwidth]{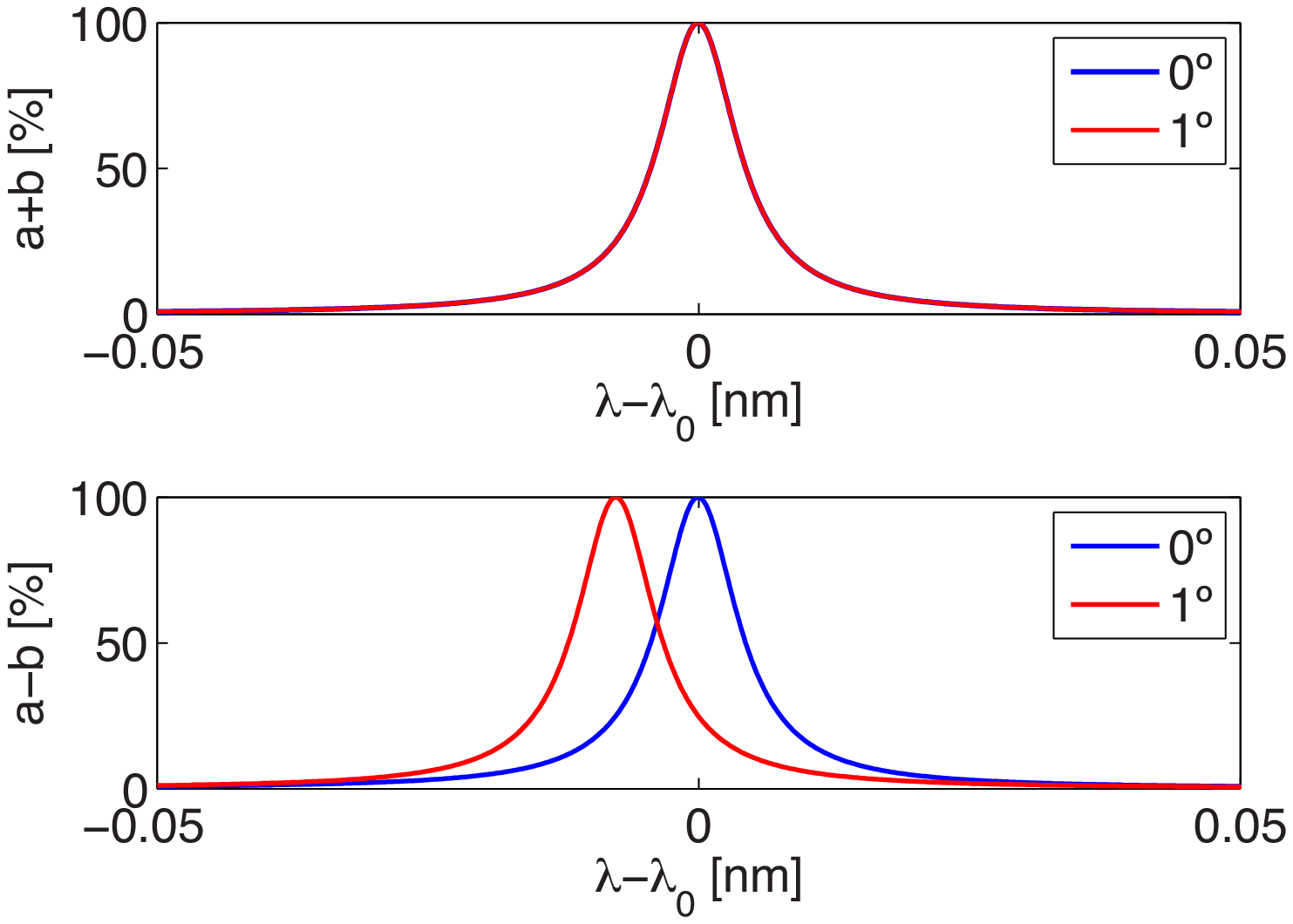}
	\end{center}
	\caption{Spectral dependence of the transmission profile of the ordinary ray ($a+b$) and of the extraordinary ray ($a-b$) for normal illumination and two different angles of the optical axis: $0\degree$ (blue solid line)  and $1\degree$ (red solid line)}
	\label{ab_theta3}
\end{figure}

The birefringence is more noticeable when varying the optical axis angle than when changing the incident angle since the dependence with $\theta_3$ is  stronger than with $\theta$ in Equation~(\ref{varphi_aprox}). 
We have also represented the ordinary ($a+b$) and extraordinary ($a-b$) spectral transmission profiles in Fig.~\ref{ab_theta3} using Eq.~(\ref{aplusb}) and Eq.~(\ref{aminusb}) for $\theta_3=0\degree$ and $1\degree$ to check both that the ordinary transmission ray does not shift to the blue, in contrast to the extraordinary ray, and that the ordinary and extraordinary profiles are symmetric. 

 \subsection{Etalon response in telecentric configuration}
 \label{tele}
In a telecentric configuration, each point of the etalon is illuminated by an identical cone of rays coming from the pupil. This means, on the one hand, that each point of the etalon receives a set of rays, each one with different angles of incidence. On the other hand, the principal plane orientation changes with the direction of each particular incident ray. We must take care of both effects. 

Consider an etalon within an optical system in a perfect telecentric configuration, where the chief ray is parallel to the optical axis over the whole field of view (FOV).  In such an ideal configuration, the etalon receives the same cone of rays across the FOV and the polarization response remains equal over the image.  Without loss of generality, we can calculate the transmitted electric field at the center of the image and extend this result to all the points of the FOV. The only dependence of the electric field with the coordinates of the pupil ($r$,$\phi$) is that of the ordinary and extraordinary retardances through the incidence angle $\theta(r)$:

\begin{equation}
\theta(r)=\arcsin{\left(\frac{r}{\sqrt{r^2+f^2}}\right)}.
\end{equation}

 Equation (\ref{Jones}) neglects any corrections in the Mueller matrix when integrating over the azimuthal direction. Rotations of the principal plane over the cone of rays must be included, though. To do so, let us first rotate the Jones matrix an angle $\phi'$. The components of the rotated Jones matrix, $\rm {\textbf H}'$ will be given by

\begin{equation}
\begin{gathered}
\H\prima_{11}=\H_{11}\cos^2\phi\prima+ \H_{22}\sin^2\phi\prima,\\
\H'_{22}=\H_{11}\sin^2\phi'+ \H_{22}\cos^2\phi',\\
\H'_{12}=\H'_{21}=(\H_{11}-\H_{22})\sin\phi'\cos\phi'.\\
\end{gathered}
\label{Hprima}
\end{equation}
The transmitted electric field shall then be calculated from the ``integrated'' Jones matrix $\Htilde\prima$, whose elements can be obtained from the Fraunhofer integral of the coefficients $\rm H_{ij}\prima$ (see Appendix \ref{app:Jones} for further details). As explained in Paper I, the resulting integrals have no easy analytical integration and shall be evaluated numerically. 

In an ideal telecentric configuration, the only dependence on the azimuthal angle is due to the rotation of the principal plane over the interval of integration. The Jones matrix elements of the telecentric configuration, $\Htildenb_{ij}\prima$, can then be cast as (Appendix \ref{app:Jones})

\begin{equation}
\begin{gathered}
\Htildeel'_{11} =\Htildeel'_{22}= \frac{1}{2}(\Htildeel_{11}+\Htildeel_{22}),\\
\Htildeel'_{12} = \Htildeel'_{21}=0,
\end{gathered}
\label{Htildeprima}
\end{equation}
where,
\begin{equation}
\begin{gathered}
\Htildeel_{11}=\int_{0}^{R_{\rm p}}\!\!\!\!\!r\H_{11}(r)\,{\rm d}r,\\
\Htildeel_{22}=\int_{0}^{R_{\rm p}}\!\!\!\!\!r\H_{22}(r)\,{\rm d}r,\\
\end{gathered}
\label{Htilde}
\end{equation}
and $R_{\rm p}$ is the radius of the pupil.
The diagonal elements of the rotated Jones matrix are therefore a linear combination of the elements of the non-rotated Jones matrix. Cross-talks between the ordinary and extraordinary components of the electric field are canceled out when integrating $\sin\phi\cos\phi$ over ($0,2\pi$). The Mueller matrix coefficients are then given by
\begin{equation}
{\rm \tilde{M}_{ij}}\prima=\frac{1}{2}{\rm Tr}\left[\sigma_i\tilde{{\Hmatrix}\prima}\sigma_{j}\tilde{{\Hmatrix}\prima}^\dagger\right].
\label{Mtilde}
\end{equation}
 Obviously, this matrix has again the form of Eq.~(\ref{M}) with coefficients $\tilde{a}\prima,\tilde{b}\prima,\tilde{c}\prima,\tilde{d}\prima$:

\begin{equation}
{\rm \tilde{\textbf{M}}}\prima=
\begin{pmatrix}
\tilde{a}\prima & \tilde{b}\prima & 0 & 0\\
\tilde{b}\prima & \tilde{a}\prima & 0 & 0\\
0 & 0 & \tilde{c}\prima & -\tilde{d}\prima\\
0 & 0 & \tilde{d}\prima& \tilde{c}\prima
\end{pmatrix},
\label{Mtele}
\end{equation}
where
\begin{equation}
\begin{gathered}
\tilde{a}\prima=\frac{1}{2}(\Htildeel_{11}\prima\Htildeel\primast_{11}+\Htildeel_{22}\prima\Htildeel\primast_{22}),\\
\tilde{b}\prima=\frac{1}{2}(\Htildeel_{11}\prima\Htildeel\primast_{11}-\Htildeel_{22}\prima\Htildeel\primast_{22}),\\
\tilde{c}\prima=\frac{1}{2}(\Htildeel_{22}\prima\Htildeel_{11}\primast+\Htildeel_{11}\prima\Htildeel_{22}\primast),\\
\tilde{d}\prima=\frac{\rm i}{2}(\Htildeel_{22}\prima\Htildeel_{11}\primast-\Htildeel_{11}\prima\Htildeel_{22}\primast).
\end{gathered}
\label{eq:abcdtilde}
\end{equation}
In this particular case, substituting Eq. (\ref{Htildeprima}),
\begin{equation}
\begin{gathered}
\tilde{a}\prima=\tilde{c}\prima=\frac{1}{4}(\Htildenb_{11}\Htildenb_{11}^*+\Htildenb_{22}\Htildenb_{22}^*+\Htildenb_{11}\Htildenb_{22}^*+\Htildenb_{22}\Htildenb_{11}^*),\\
\tilde{b}\prima=\tilde{d}\prima=0.
\end{gathered}
\label{abcdprima}
\end{equation}

Therefore, the Mueller matrix is diagonal and no cross-talk appears between the different spectral profiles of the Stokes components if telecentricity is exact. 

However, perfect telecentricity can only be reached ideally. In a normal scenario, there is a dependence with the azimuthal angle on the Jones matrix even if changes of orientation of the principal plane over the azimuthal angle $\phi$ were not considered, since the symmetry of the problem is broken. In an imperfect telecentric configuration, the chief ray direction deviates from normal incidence on the etalon over the FOV and so does the polarimetric response of the etalon, which is now expected to be spectrally asymmetric. Equations (\ref{Htildeprima}) and (\ref{Mtele}) cannot be applied. Actually $a\prima$ is no longer equal to $c\prima$ and $b\prima$ and $d\prima$ become different from zero. Furthermore, the Jones matrix off-diagonal elements are, in general, different from zero. The Mueller matrix elements should be calculated from Eq.~(\ref{Mtilde}) with the coefficients $\rm H_{ij}\prima$ obtained following Appendix \ref{app:Jones}.

\begin{figure}[t]
 	\centering
 	\includegraphics[width=0.5\textwidth]{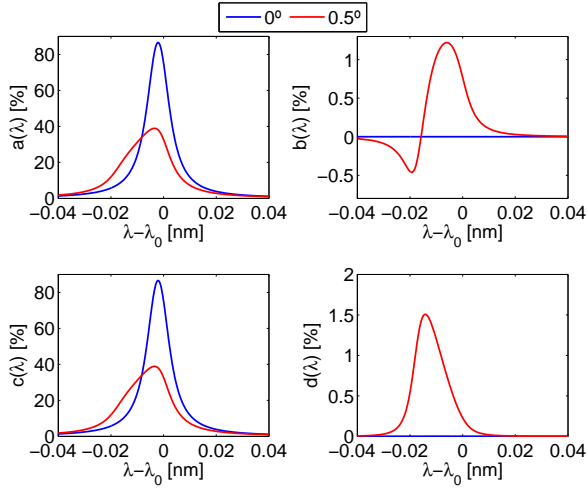}
 	\caption{Variation of the $\tilde{a}\prima, \tilde{b}\prima, \tilde{c}\prima$ and $\tilde{d}\prima$ coefficients of the Mueller matrix of the etalon as a function of the wavelength for both perfect telecentrism (blue solid line) and imperfect telecentrism with a deviation of the chief ray of $0\fdeg5$ (red solid line). A beam f-number of $60$ has been employed.}
 	\label{abcd_tilde}
 \end{figure}

Figure \ref{abcd_tilde} represents the spectral response of the Mueller matrix elements as a function of $\lambda$ for an optical system with $f/60$. Both a perfect telecentric configuration (chief ray at $0\degree$) and an imperfect telecentrism in which the chief ray is deviated $0\fdeg5$ have been considered.
We only show the $\tilde{a}\prima, \tilde{b}\prima, \tilde{c}\prima,$ and $\tilde{d}\prima$ components of the Mueller matrix since we have observed in our numerical experiments that other off-diagonal elements in the Jones matrix are several orders of magnitude below the diagonal terms. This implies that, in practice, $\Htilde\prima$ can be considered as diagonal and only the coefficients $\tilde{a}\prima, \tilde{b}\prima, \tilde{c}\prima,$ and $\tilde{d}\prima$ need to be calculated. 

First to notice is that the profiles are blue-shifted, as in the collimated configuration. We also see how $\tilde{a}\prima$ and $ \tilde{c}\prima$ profiles for imperfect telecentrism are broader. Their peak values have decreased from about $\sim 90\%$ at $0\degree$ to $\sim 40\%$ at $0\fdeg5$  due to the mentioned widening. These two effects are more important for shorter f-numbers because of the larger incidence angles (Paper I). 
 
Remarkably, the four matrix elements have a clear asymmetric spectral dependence at $0\fdeg5$. The maximum values of $\tilde{b}\prima$ and $\tilde{d}\prima$ are $\sim 1\%$ and$\sim 1.5\%$ respectively. 
These terms are responsible for the cross-talk among the Stokes parameters. Note that these large asymmetries in the spectral profile are not exclusive for birefringent etalons, since they also appeared in Paper I, where the isotropic case was studied. At $0\degree$ there is no cross-talk and $\tilde{a}\prima(\lambda)=\tilde{c}\prima(\lambda)$,  as expected from Equation (\ref{abcdprima}). Although not noticeable in this figure, the loss of symmetry in an imperfect telecentrism implies that $\tilde{a}\prima(\lambda)\neq\tilde{c}\prima(\lambda)$ at $0\fdeg5$, as explained before.
 
\begin{figure*}[t]
	\plottwo{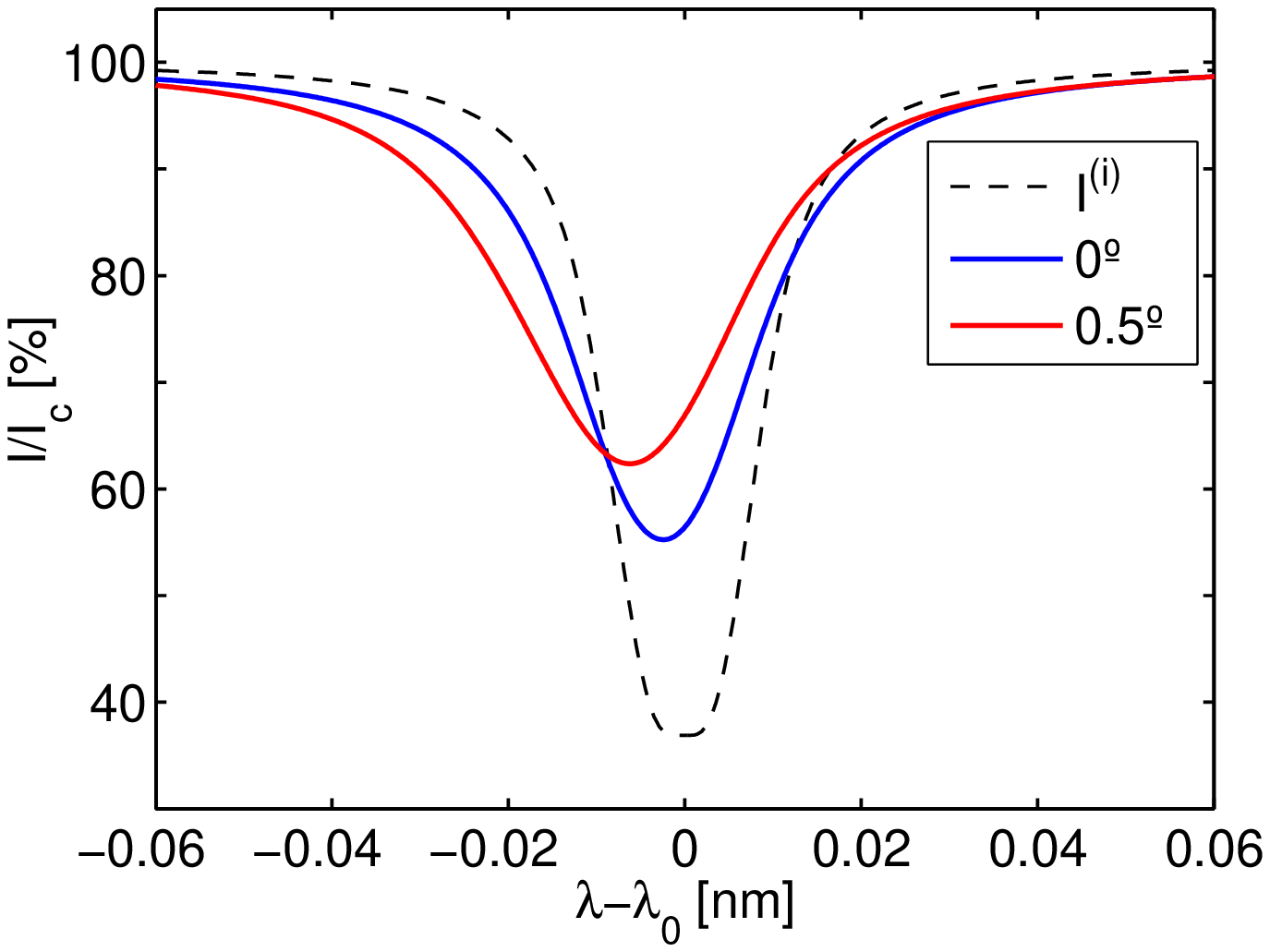}{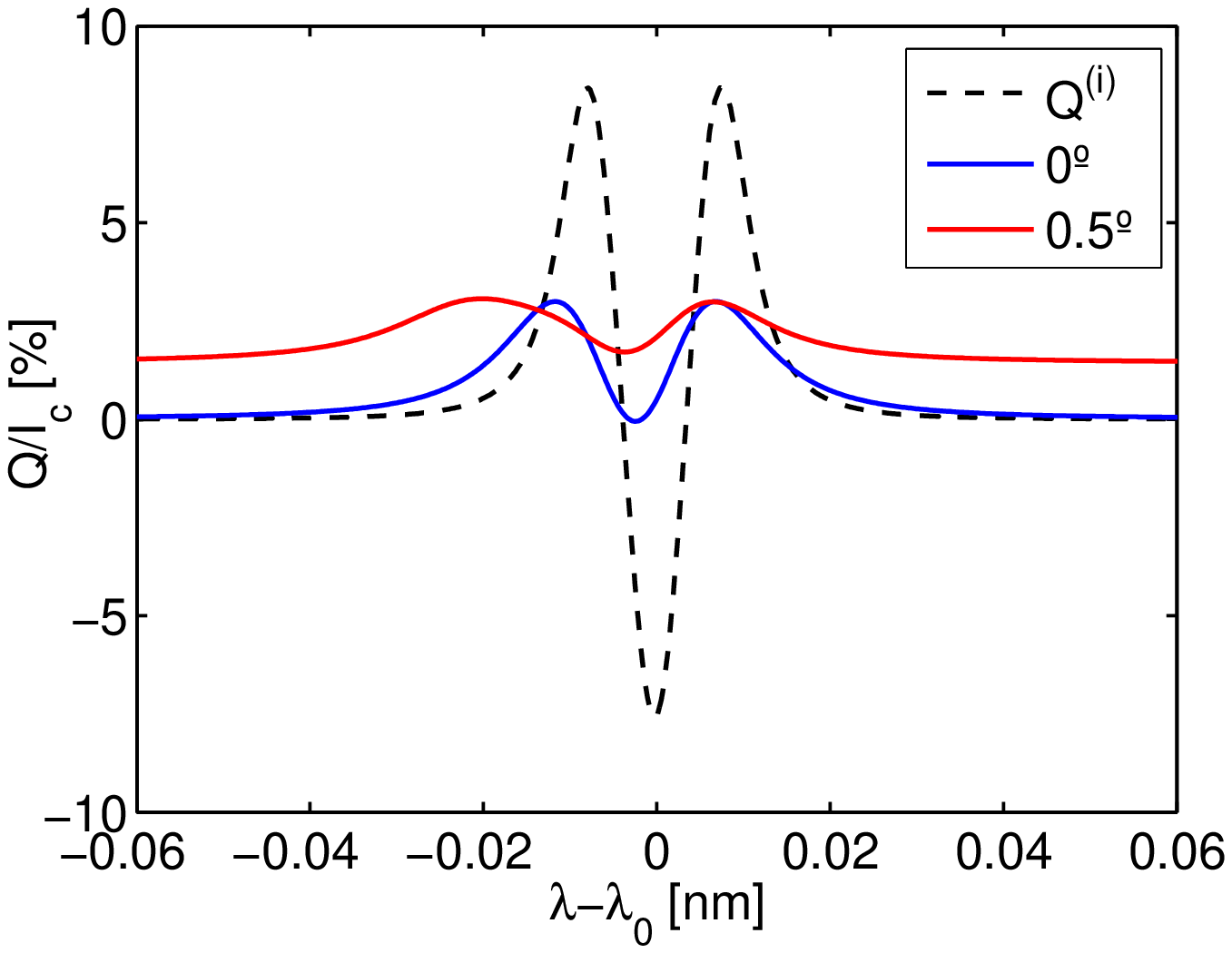}
	\plottwo{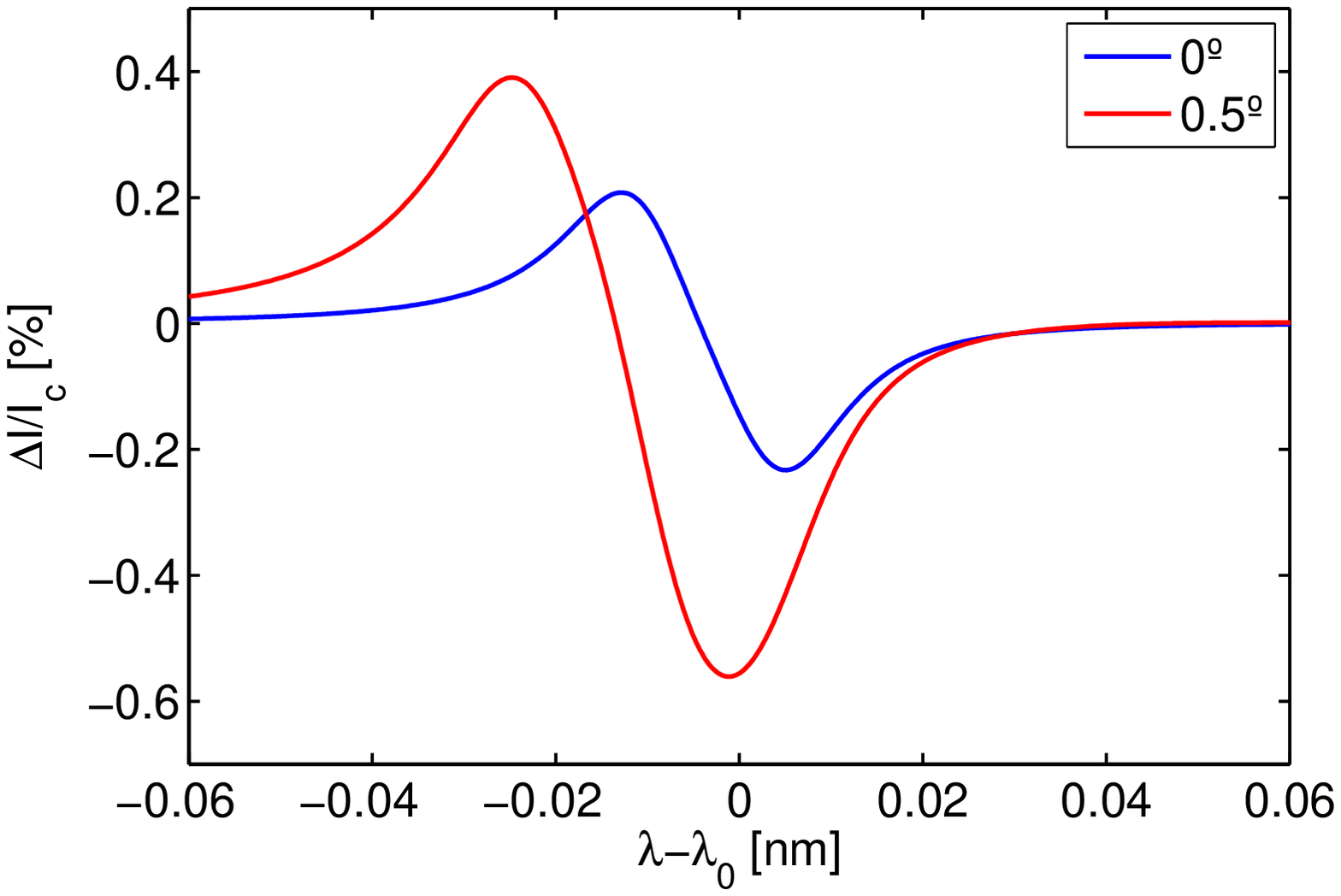}{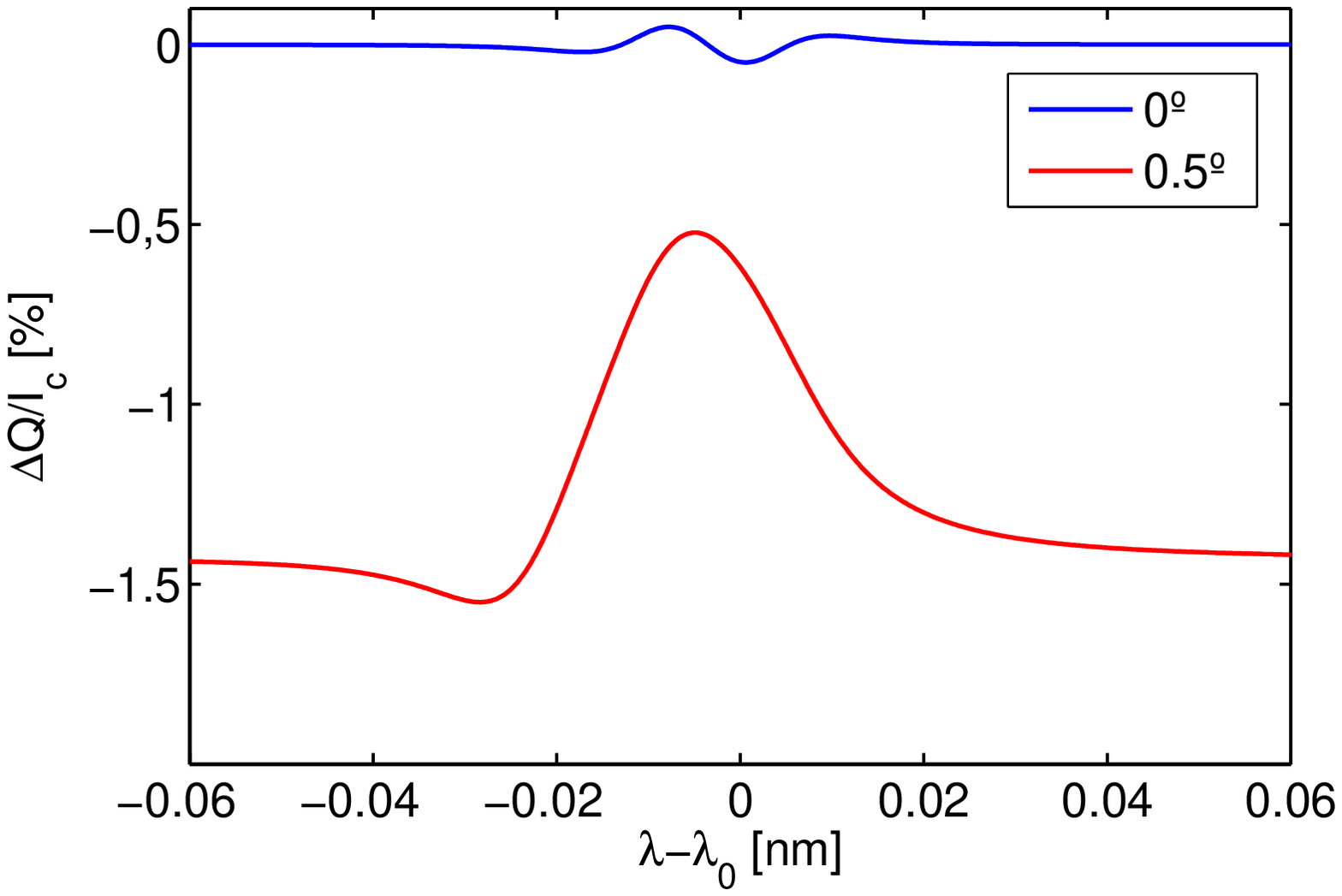}
	\caption{Observed Stokes $I$ and $Q$ profiles (left and right panels, respectively) when illuminating the etalon in telecentric configuration with f/60 with the orientation of the chief ray at $0\degree$ (blue solid line) and at $0\fdeg5$ (red solid line). Black dashed stands for the synthetic input Stokes $I$ and $Q$ profiles. The cross-talks induced in Stokes $I$ and Stokes $Q$ are shown in the bottom panels.}
	\label{IQ_tilde}
\end{figure*}
Figure~\ref{IQ_tilde} shows the observed Stokes $I$ and $Q$ spectral profiles when illuminating the etalon with the same synthetic profile as in Sec. \ref{Oblique}, and using a telecentric configuration with f/60 as well.
We can see the displacement towards the blue produced by the effect of the different incidence angles. The profiles also broaden due to the effect of the convolution with the Mueller matrix of the etalon (Eqs. \ref{Iconv} and \ref{Qconv}) and become asymmetric. Moreover, an artificial continuous signal in the measured $Q$ at $|\lambda-\lambda_0|>0.03$~nm appears due to the cross-talk introduced from the continuous part of $I$. In order to estimate the induced artificial signals due to the birefringence of the etalon, we have also plotted $\Delta I=I^{(t)}-I^{(t)}_{\rm nb}$ and $\Delta Q=Q^{(t)}-Q^{(t)}_{\rm nb}$, where $I^{(t)}_{\rm nb}$ and $Q^{(t)}_{\rm nb}$ are the transmitted $I$ and $Q$ components of the Stokes vector for a non-birefringent etalon with refraction index $n_o$. The absolute maximum cross-talk goes from $\sim 0.2\%$ and $\sim 0.05\%$ at $0\degree$ to $\sim 0.65\%$ and $\sim 1.75\%$ at $0\fdeg5$ in $I$ and $Q$ respectively.

\section{Imaging response to monochromatic plane waves} \label{sec:PSF}
As discussed in Paper I, space invariance is not preserved in neither the collimated nor the (imperfect) telecentric case. We cannot speak, then, of a PSF that can be convolved with the object brightness distribution when studying the response of the etalon. Instead, we have to integrate the object with a \emph{local} PSF. On the other hand, since the object brightness usually varies with wavelength, the response of the Fabry-P\'erot depends on the object itself. We need then to integrate spectrally the monochromatic response of the instrument (Eqs. [61] and [62] of Paper I).
Moreover, orthogonal components of the electric field are, in general, modified in a different way when traversing through the etalon. We expect therefore the response to vary with the incident polarization as well.

The \emph{local} PSF, $\PSF$, is defined as the ratio
\begin{equation}
{\rm \PSF}=\frac{\tilde{\textbf {E}}^{(\text{t}){\ast}}\tilde{\textbf {E}}^{(\text{t})}}{\textbf{E}^{(\text{i})\ast} {\textbf {E}}^{(\text{i})}},
\label{PSF_E}
\end{equation}
where $\Ei$ is the electric field of the incident plane wave and $\tilde{\textbf {E}}^{(\text{t})}$ is the image plane electric field, related to the incident ordinary and extraordinary rays by
\begin{equation}
\begin{gathered}
\Etilde_x=\Htildeel_{11}\prima\Ei_o+\Htildenb_{12}\prima\Ei_e,\\
\Etilde_y=\Htildeel_{21}\prima\Ei_o+\Htildenb_{22}\prima\Ei_e.
\end{gathered}
\label{Erotations}
\end{equation}
In a similar way to Section \ref{tele}, coefficients $\Htildeel_{ij}\prima$ are calculated from the Fraunhofer integrals (Appendix \ref{app:Jones}) of the elements of the ``rotated'' Jones matrix, $\rm H_{ij}\prima$, and depend on the image plane coordinates $(\xi,\eta)$, on the chief ray coordinate in the image plane $(\xi_0,\eta_0)$, and on the wavelength. We do not explicit these dependences in the equations that follow for simplicity. Note that we do not restrict ourselves now to the center of the image, unlike in Section \ref{tele}, since we are interested not only on the transmission profiles of the Stokes vector but on the consequences of diffraction effects due to the limited aperture of the system.

Even if we neglect crossed terms in the Jones matrix, the response of the etalon is determined by the polarization of the incident light, since the diagonal terms of the Jones matrix are different. This statement is valid for both collimated and telecentric mounts. For isotropic media, since $\Htildenb\prima_{12}=\Htildenb\prima_{21}=0$ and $\Htildenb\prima_{11}\Htildenb_{11}^{\prime\ast}=\Htildenb\prima_{22}\Htildenb_{22}^{\prime\ast}$, we recover the result for $\PSF$ presented in Paper I.

Equation (\ref{Erotations}) is written in terms of the ordinary and extraordinary electric field components. It may be more useful to find the relation of $\PSF$ with the incident Stokes parameters, though. This is as easy as obtaining the Mueller matrix through Eq.~(\ref{Mtilde}) and noticing that $\tilde{\textbf {E}}^{(\text{t}){\ast}}\Etilde$ represents the first component of the transmitted Stokes vector. Consequently, substituting in Eq. (\ref{PSF_E}),
\begin{equation}
\PSF=\Mijtilde_{11}+\Mijtilde_{12}\frac{Q^{(i)}}{I^{(i)}}+\Mijtilde_{13}\frac{U^{(i)}}{I^{(i)}}+\Mijtilde_{14}\frac{V^{(i)}}{I^{(i)}}.
\label{eq:PSF_stokes}
\end{equation}

Again, we can see that $\PSF$ depends in general on the polarization of the incident light. Although the expressions presented in this section are valid for both collimated and telecentric illumination, differences between both cases are obviously expected to arise, so the need to study them separately.

\subsection{Collimated configuration}
In the collimated configuration, there is a one-to-one mapping between the incidence angle of the rays on the etalon and their position on the image plane. As only the incidence angles are of interest, the location of the rays on the pupil is irrelevant and the Fraunhofer integrals are proportional to that of a circular aperture with the same radius, similarly to the isotropic case. According to Appendix \ref{app:Jones}, the Jones matrix terms are actually given by
\begin{equation}
\begin{gathered}
\Htildeel\prima_{11}=\left(\H_{11}\cos^2\phi_0+ \H_{22}\sin^2\phi_0\right)\frac{2J_1(z)}{z},\\
\Htildeel'_{22}=\left(\H_{11}\sin^2\phi_0+ \H_{22}\cos^2\phi_0\right)\frac{2J_1(z)}{z},\\
\Htildeel'_{12}=\Htildeel'_{21}=\left(\H_{11}-\H_{22}\right)\sin\phi_0\cos\phi_0\frac{2J_1(z)}{z},\\
\end{gathered}
\label{Hprima_coll}
\end{equation}	
where the variable $z$ and $J_1$ are defined in Paper I, and
 $\phi_0$ is the azimuthal orientation of the principal plane of the etalon with respect to the $+Q$ direction of the  reference frame chosen to describe the Stokes vector (i.e., the azimuthal angle in Figure~\ref{xyz}). 
Note that off-diagonal terms cannot be neglected unless $\phi_0=0$. Thus, $\PSF$ depends on the four Stokes parameters and varies over the image plane due to both the birefringence of the etalon and the re-orientation of the principal plane with the incident ray direction. In fact, for the same radial position on the image plane, $\PSF$ changes because of the different orientations of the principal plane. A decrease of the intensity is also expected towards the edges of the image, as explained in Paper I. 

We can only set $\phi_0=0$ for ray directions parallel to the optical axis. Assuming the optical axis is perpendicular to the surfaces of the etalon, this occurs at normal illumination of the pupil. For this particular case, the Mueller matrix has the form of Eq.~(\ref{Mtele}) and, using Eqs.~(\ref{eq:PSF_stokes}) and (\ref{Hprima_coll}), an analytical expression for $\PSF$  can be found:

\begin{equation}
\begin{gathered}
{\rm \PSF}=\frac{1}{2}\Airy\left[g_o+g_e+(g_o-g_e)\frac{Q^{(i)}}{I^{(i)}}\right],
\label{PSFc2}
\end{gathered}
\end{equation}
where $g_o\equiv\H_{11}\H_{11}^\ast$ and $g_e\equiv\H_{22}\H_{22}^\ast$ are the transmission profiles of the ordinary and extraordinary rays for normal illumination of the etalon. This expression illustrates the polarimetric dependence of $\PSF$ for the collimated configuration and its proportionality to that of an ideal circular aperture. Notice that, since crossed terms in the Jones matrix are zero in this case, $\Mijtilde_{13}$ and $\Mijtilde_{14}$ are also null, and the dependence with Stokes components $U$ and $V$ disappears. For pupil incidence angles different from zero, expressions are much more involved and an analytical expression for $S$ cannot easily be obtained. 

\subsection{Telecentric configuration}
For telecentric illumination of the etalon, the retardance is related to the pupil coordinates of the incident rays, unlike for the collimated case. The proportionality with the response of a circular aperture disappears then, as occurred in the isotropic case, and the Jones matrix elements of Eq.~(\ref{Erotations}) must be evaluated numerically.
 
The response $\PSF$, as for the collimated case, depends on the polarization state of the incident light even for perfect telecentrism. This is because $\Htildenb_{11}'\neq \Htildenb_{22}'$ and $\Htildenb_{12}'=\Htildenb_{21}'\neq 0$ in general, as explained in Appendix \ref{app:Jones}. Let us consider two simple cases, namely, $\Ei_e=0$ and $\Ei_o=0$. For the  case $\Ei_e=0$, according to Eq.~(\ref{Erotations}), the PSF follows the expression
	\begin{equation}
	{\rm \PSF}=\Htildenb_{11}'\Htildenb_{11}'^*+\Htildenb_{21}'\Htildenb_{21}'^*.
	\label{PSFEe0}
	\end{equation} 
	For the case $\Ei_o=0$, the PSF is described by
	\begin{equation}
	{\rm \PSF}=\Htildenb_{22}'\Htildenb_{22}'^*+\Htildenb_{12}'\Htildenb_{12}'^*,
	\end{equation} 
which is different from Eq.~(\ref{PSFEe0}) even if we ignore the cross-talk term (second term of the equations). Cross-talks can be neglected in practice for the telecentric configuration, as discussed in Section \ref{tele}. Therefore, the third and fourth Mueller matrix terms of Eq.~(\ref{eq:PSF_stokes}) vanish, as for normal illumination in collimated etalons, and the response depends only on $I$ and $Q$ Stokes components (as well as on the birefringence of the etalon). Interestingly, the peak of $\PSF$ is just the transmission profile, $\tilde{a}\prima$, (Eq.~\ref{abcdprima}) which is not affected by the incident polarization state of light.

Obviously, if the chief ray is not perpendicular to the etalon surfaces (imperfect telecentrism) the same arguments can be applied. Moreover, other effects explained in Paper I will appear. Essentially, $\PSF$ becomes asymmetric and vary from point to point. If telecentrism is perfect, although polarization-dependent, $\PSF$ remains the same all over the FOV by definition.

\begin{figure}
	\plotone{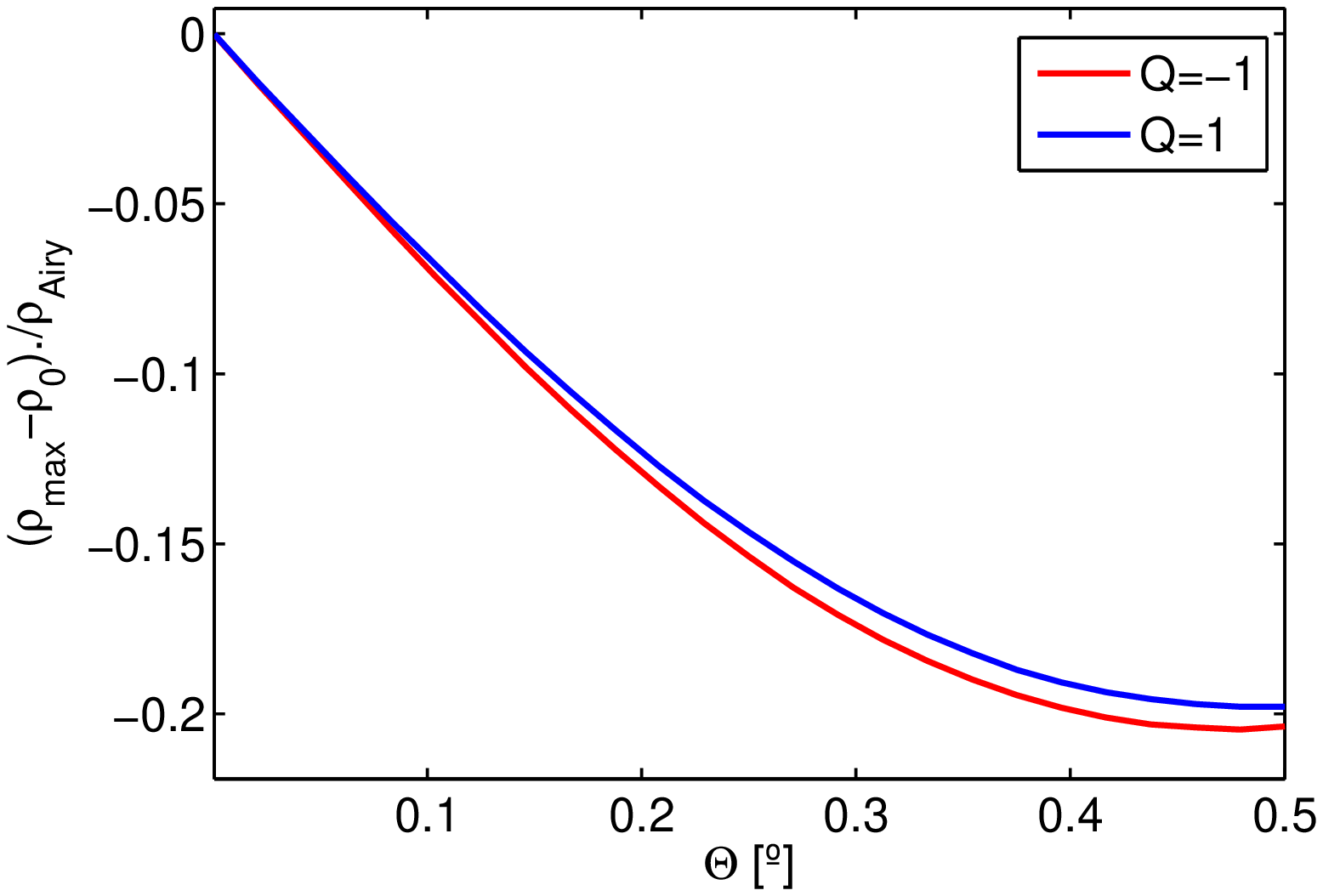}
	\plotone{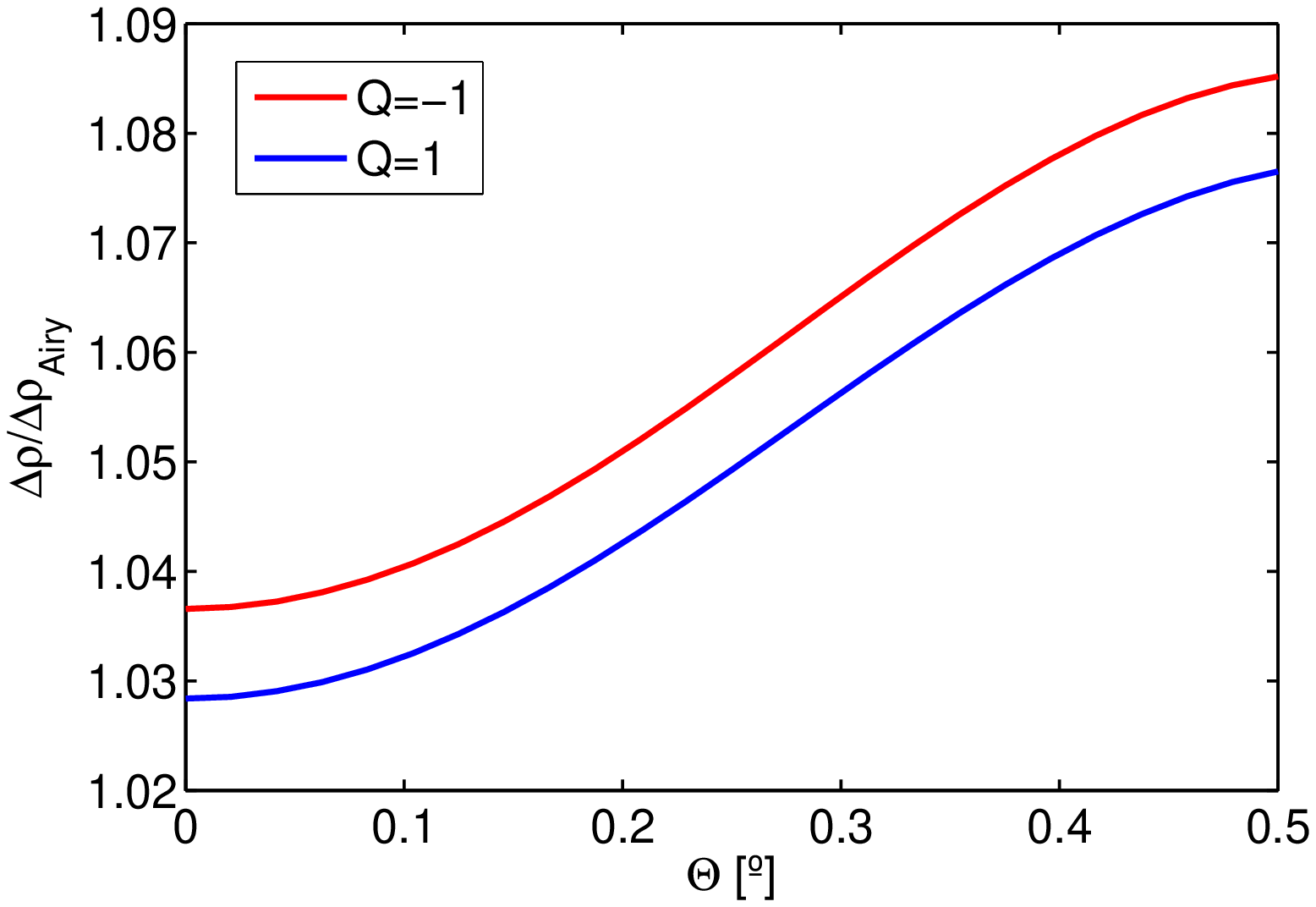}
	\caption{Spatial shift (top) of the peak of the PSF with respect to the Airy disk and FWHM of the PSF normalized to that of the Airy disk (bottom) as we move across the field of view in the image plane ($X$-axis). The plot include the result for orthogonal polarization beams $Q=1$ (blue) and $Q=-1$ (red). A telecentric beam with $f/60$ has been employed.}
	\label{PSF-dep}
\end{figure}

To evaluate how $\PSF$ varies with the polarization of the incident light beam, we have calculated its width and its peak position for $Q=\pm 1$ states of polarization. We study their behavior with the degree of telecentrism by varying the chief ray angle, $\Theta$, from $0\degree$ (ideal telecentrism) to $0\fdeg5$. Figure \ref{PSF-dep} shows the results obtained for an $f/60$ beam. The $X$ axis of both top and bottom figures indicates the angle that the chief ray forms with the optical axis.

 The top figure represents the shift of the PSF peak with respect to the position of the peak for the ideal diffraction-limited case (Airy disk) with $\Theta$ and for the two orthogonal polarizations. The results have been normalized by the radius of the collimated case. It can be seen clearly that the shift is different for orthogonal polarizations, meaning that $\PSF$ depends on the input beam polarization state. Deviations between orthogonal states would be larger for smaller $f$ ratios. 

The bottom figure shows the width of $\PSF$ normalized to that corresponding for the ideal diffraction-limited case. Notice that apart from an offset between the two curves, they depend slightly different with $\Theta$. The offset indicates that $\PSF$ is polarization-dependent even for perfect telecentrism (i.e., when $\Theta=0$), as explained before in the text.

\section{Comments on the birefringent effects in solar instrumentation}
\label{sec:comments}
The polarimetric effects described in earlier sections have an impact on the incident Stokes vector. Indeed, off-diagonal terms in the etalon Mueller matrix introduce cross-talks between the Stokes parameters that could deteriorate the measurements carried out by solar magnetographs. However, there are other factors that should also be considered for a proper evaluation of the spurious signals emerging in such instruments.

First, we need to take into account the combined response of the polarimeter and the etalon because both modify the polarization state of light. The final Mueller matrix of the instrument depends, then, on the relative position of the etalon with respect to the polarimeter. Usually, the Fabry-P\'erot is located either between the modulator and the analyzer or behind it. When it is located after the analyzer, further cross-talks induced by the etalon are prevented. The reason for this is that the etalon is illuminated with linearly polarized light. If placed between the modulator and the analyzer, then the effect on the final Mueller matrix changes for each particular modulation of the signal.

Second, observations are not strictly monochromatic but quasi-monochromatic. Spectral integration of the Mueller coefficients decreases the magnitude of cross-talk terms, specially for $b$ and $\tilde{b}\prima$ since they change their sign along the spectral profile (Figures \ref{abcd1d} and \ref{abcd2d}). 

Modulation of the signal and the quasi-monochromatic nature of the observations reduce the cross-talk induced by the etalon Mueller matrix. These aspects will be addressed in the next work of this series of papers. 

The calculations presented in previous sections represent a worst-case scenario.  Let us consider two examples of instruments based on birefringent etalons: SO/PHI \citep{ref:sophi} and IMaX \citep{imax}. The former is illuminated with a telecentric beam, whereas the second is mounted on a collimated configuration.  For SO/PHI, the degree of telecentrism is kept below $0\fdeg 23$ in a $\sim f/60$ mount. In addition, its etalon is located after the analyzer. In the IMaX instrument, incidence angles are below $0\fdeg44$ and the Fabry-P\'erot is placed between the modulator and the analyzer. Deviations from normal illumination in SO/PHI and IMaX  are lower than half the maximum angle employed in Figures \ref{abcd_tilde} and \ref{abcd1d}. Moreover, deviations of the optical axis from the nominal one have only been observed to appear after the application in the laboratory of very intense electric fields and disappear after a certain interval of time. These deviations are distributed in small compact regions or \emph{local domains} that cover a small fraction of the clear aperture. If these electric fields are not reached during operation, the harming effects can be considered negligible.

\section{Summary and conclusions} \label{sec:conclusions}

A general theory that considers the polarimetric response of anisotropic (uniaxial) crystalline etalons has been presented in this work. 
We have obtained an expression of the Mueller matrix that describes the polarimetric behavior of uniaxial crystalline etalons and we have concluded that they can be described as a combination of an ideal mirror and a retarder, both strongly spectrally modulated. We have shown that the Mueller matrix of the etalon in a collimated configuration depends only on four elements that vary spectrally, with the direction of the incident rays and on the orientation of the optical axis. A careful choice of the reference frame depending on the orientation of the principal plane is also needed. 

We have also deduced an analytical expression for the birefringence induced in uniaxial crystalline Fabry-P\'erot etalons that takes into account both the direction of the incident rays and the orientation of the optical axis.
 By numerical experimentation, we have studied the effect of (1) oblique illumination in $Z$-cut etalons; (2) misalignments of the optical axis at normal illumination; and (3)  locating the etalon in a telecentric configuration. We have considered the influence of illuminating with different f-numbers in the latter. 

For the first case, we have evaluated the spectral dependence of the coefficients of the Mueller matrix with the angle of the incident light. 
We have shown that, with the parameters of a commercial etalon, the cross-talk between $I$ and $Q$ is about 10$\%$ at 1$\degree$ and 30$\%$ between $U$ and $V$. For the second case, we have showed that the same deviations of the optical axis introduce larger artificial signals between the Stokes parameter (40$\%$ and 60$\%$ between $I$ and $Q$ and $U$ and $V$ at 1$\degree$).
We have also evaluated the spectral transmission of a synthetic Stokes profile when traversing through the etalon for different incident angles. Asymmetries are induced in this case in the observed profiles due to the presence of cross-talk terms in the Mueller matrix, thus introducing spurious signals. 

We have shown that in a perfect telecentric configuration, the Mueller matrix is diagonal and no cross-talk appear between the different Stokes components. For an imperfect telecentric beam, the Mueller matrix is not diagonal anymore, although it still keeps the  $abcd$ form in practice, and the spectral profiles of the Mueller matrix elements become asymmetric.
We have studied the spectral profiles of the Mueller matrix coefficients and the degradation produced on a spectral artificial Stokes profile and we have estimated the cross-talks produced in this configuration. Because of the birefringence of the etalon, artificial signals appear on the observed profile compared to the isotropic case, apart from the known broadening and blueshift effects.

A general method for obtaining the imaging response in crystalline Fabry-P\'erots for both collimated and telecentric configurations has been developed. It has been shown that the response of the etalon is related in general to the polarization of the incident light, as well as to its birefringence. We have addressed the problem from two different points of view: by using the Jones formalism and by employing the Mueller matrix method. Both of them are equivalent. The advantage of the second is that it let us express the response directly as a function of the input Stokes parameters. 
	
We have demonstrated that in a collimated setup the \emph{local} PSF is modified with respect to the ideal PSF by a transmission factor that varies across the image plane both radially and azimuthally due to the correspondent rotations of the principal plane with the ray direction (Eq.~\ref{Hprima_coll}). At the origin, the response is equal to the irradiance distribution of a circular unaberrated pupil modulated by a transmission factor that depends on the birefringence of the etalon and on the $I$ and $Q$ Stokes components that traverse through the etalon. In a perfect telecentric configuration the PSF also depends on the induced birefringence of the etalon and on the incident polarization state of light (namely, on $I$ and $Q$ again), although its peak transmission is polarization independent and its shape remains the same across the image plane. In imperfect telecentrism, an asymmetry and a variation of the response over the detector are also introduced. We have evaluated the spatial shift of the response for two orthogonal states of polarization with the degree of telecentrism, as well as its FWHM. We have shown that the local PSF peak and FWHM change different with the chief ray angle for each polarization. The FWHM depends on the polarization of the incident light even for perfect telecentrism. The numerical results obtained are in agreement with our analytical argumentation.

\begin{acknowledgements}
This work has been supported by Spanish Ministry of Economy and Competitiveness through projects ESP2014-56169-C6-1-R and ESP-2016-77548-C5-1-R. The authors acknowledge financial support from the State Agency for Research of the Spanish MCIU through the "Center of Excellence Severo Ochoa" award for the Instituto de Astrof\'isica de Andaluc\'ia (SEV-2017-0709). DOS also acknowledges financial support through the Ram\'on y Cajal fellowship.
\end{acknowledgements}

\appendix
\section{A: Exact expression of the electric field at the focal plane}
\label{app:Jones}
The electric field of an electromagnetic wave at the focal plane of an optical instrument, $\Etilde(\xi,\eta;\xi_0,\eta_0;\lambda)$, is given by the Fraunhofer integral of the incident electric field at the pupil, $\Ei$. We have remarked the dependence of the electric field with the coordinates of the focal plane ($\xi,\eta$); the chief ray position at the focal plane ($\xi_0,\eta_0$); and the wavelength, since they are variables of interest for the calculation of the spectral transmission profile and of  the monochromatic imaging response.  We omit these explicit dependences from this point on.

If we choose radial coordinates $(r,\phi)$ to describe the pupil coordinates, we can write
\begin{equation}
\Etilde=\int_{0}^{R_{\rm p}}\!\!\!\!\!\int_{0}^{2\pi}\!\!\!\!\!\!r\Et(r,\phi,\lambda) {\rm e}^{-{\rm i} kr(\alpha \cos\phi + \beta \sin\phi)}\,{\rm d}r \, {\rm d}\phi,
\end{equation}
where $k$ is the wavelength vector of the incident wavefront, $\alpha\equiv(\xi-\xi_0)/f$ and $\beta\equiv(\eta-\eta_0)/f$ are the cosine directors (not to be confused with the angles of Fig.~\ref{xyz}), and $R_{\rm p}$ is the radius of the pupil.

Following the Jones formalism, we can also write
\begin{equation}
\Etilde\equiv\begin{pmatrix}
\Etilde_x\\
\Etilde_y
\end{pmatrix}
=\Htilde\prima
\begin{pmatrix}
\Ei_o\\
\Ei_e
\end{pmatrix},
\end{equation}
where the coefficients of $\Htilde\prima$ can be calculated from Eq.~(\ref{Hprima}) after integration:
\begin{equation}
\begin{gathered}
\Htildeel\prima_{11}=\int_{0}^{R_{\rm p}}\!\!\!\!\!\int_{0}^{2\pi}\!\!\!\!\!\!r\left[\H_{11}(r,\phi,\lambda)\cos^2\phi\prima+ \H_{22}(r,\phi,\lambda)\sin^2\phi\prima\right]{\rm e}^{-{\rm i} kr(\alpha \cos\phi + \beta \sin\phi)} \, {\rm d}r \, {\rm d}\phi,\\
\Htildeel'_{22}=\int_{0}^{R_{\rm p}}\!\!\!\!\!\int_{0}^{2\pi}\!\!\!\!\!\!r\left[\H_{11}(r,\phi,\lambda)\sin^2\phi\prima+ \H_{22}(r,\phi,\lambda)\cos^2\phi\prima\right]{\rm e}^{-{\rm i} kr(\alpha \cos\phi + \beta \sin\phi)} \, {\rm d}r \, {\rm d}\phi,\\
\Htildeel'_{12}=\Htildeel'_{21}=\int_{0}^{R_{\rm p}}\!\!\!\!\!\int_{0}^{2\pi}\!\!\!\!\!\!r\left[\H_{11}(r,\phi,\lambda)-\H_{22}(r,\phi,\lambda)\right]\sin\phi\prima\cos\phi\prima{\rm e}^{-{\rm i} kr(\alpha \cos\phi + \beta \sin\phi)} \, {\rm d}r \, {\rm d}\phi,\\
\end{gathered}
\label{exactH}
\end{equation}
where $\phi'$ is the azimuthal angle of the principal plane with respect to the $+Q$ direction of the reference frame chosen to describe the Stokes parameters (Fig. \ref{xyz}). 
 The coefficients of the Jones matrix $\Hmatrix$ are given by Eqs.~(\ref{Eo}), (\ref{Ee}) and  (\ref{Jones}). 
Note that this expression considers the rotations of the principal plane of the etalon with the ray direction vector within the etalon. 
The dependence of the Jones matrix elements with the pupil coordinates is entirely given by that of retardances $\delta_o$ and $\delta_e$ through the incidence angles and depend on the optical configuration.

\subsection{Collimated configuration}
For collimated setups, the incidence angle is given by Eq.~[53] from Paper I:
\begin{equation}
\theta=\cos^{-1}\left(\frac{f}{\sqrt{\xi^2+\eta^2+f^2}}\right),
\label{delta_coll}
\end{equation}
which does not depend on the pupil coordinates. Therefore $\H_{11}\neq\H_{11}(r,\phi)$ and $\H_{22}\neq\H_{22}(r,\phi)$, and we can cast Eq. (\ref{exactH}) as
\begin{equation}
\begin{gathered}
\Htildeel\prima_{11}=\left(\H_{11}\cos^2\phi_0+ \H_{22}\sin^2\phi_0\right)\int_{0}^{R_{\rm p}}\!\!\!\!\!\int_{0}^{2\pi}\!\!\!\!\!\!r{\rm e}^{-{\rm i} kr(\alpha \cos\phi + \beta \sin\phi)}\,{\rm d}r \,{\rm d}\phi=\left(\H_{11}\cos^2\phi_0+ \H_{22}\sin^2\phi_0\right)\frac{2J_1(z)}{z},\\
\Htildeel'_{22}=\left(\H_{11}\sin^2\phi_0+ \H_{22}\cos^2\phi_0\right)\int_{0}^{R_{\rm p}}\!\!\!\!\!\int_{0}^{2\pi}\!\!\!\!\!\!r{\rm e}^{-{\rm i} kr(\alpha \cos\phi + \beta \sin\phi)}\,{\rm d}r \,{\rm d}\phi=\left(\H_{11}\sin^2\phi_0+ \H_{22}\cos^2\phi_0\right)\frac{2J_1(z)}{z},\\
\Htildeel'_{12}=\Htildeel'_{21}=\left(\H_{11}-\H_{22}\right)\sin\phi_0\cos\phi_0\int_{0}^{R_{\rm p}}\!\!\!\!\!\int_{0}^{2\pi}\!\!\!\!\!\!r{\rm e}^{-{\rm i} kr(\alpha \cos\phi + \beta \sin\phi)}\,{\rm d}r \,{\rm d}\phi=\left(\H_{11}-\H_{22}\right)\sin\phi_0\cos\phi_0\frac{2J_1(z)}{z},\\
\end{gathered}
\end{equation}
where we denote $\phi_0$ instead of $\phi\prima$ to describe the azimuthal angle of the principal plane. This is to emphasize that $\phi_0$ does not depend on the pupil coordinates and can be taken out of the integral, since the the principal plane only changes in this case with the orientation of the incident rays, but not with their location on the pupil. The parameter $z$ is given by

\begin{equation}
z=\frac{2\pi}{\lambda}R_{\rm pup}\frac{\sqrt{\xi^2+\eta^2}}{f},
\end{equation}
and $J_1$ is the first-order Bessel function.
Whenever $\phi_0=0$ (as we can set for normal illumination of the pupil if $\theta_3=0$), the Jones matrix coefficients are greatly simplified:
\begin{equation}
\begin{gathered}
\Htildeel\prima_{11}=\H_{11}\frac{2J_1(z)}{z},\\
\Htildeel'_{22}= \H_{22}\frac{2J_1(z)}{z},\\
\Htildeel'_{12}=\Htildeel'_{21}=0.\\
\end{gathered}
\end{equation}

\subsection{Telecentric configuration} 
Unlike for the collimated configuration, a relation exists in telecentric setups between the incidence angle in the etalon and the coordinates of the pupil of the incident ray. This is described in Eq. [59] of paper I. Using radial coordinates this expression can be re-written as

\begin{equation}
\theta=\cos^{-1}\left(\frac{f}{\sqrt{(r\cos\phi-\xi)^2+(r\sin\phi-\eta)^2+f^2}}\right),
\end{equation}
and no simplification of Eq.~(\ref{exactH}) can be done in general. Only if we focus on the origin ($\xi=\eta=0$) the azimuthal dependence of $\Htilde\prima$ disappears, since
\begin{equation}
\theta=\cos^{-1}\left(\frac{f}{\sqrt{r^2+f^2}}\right).
\end{equation}
Note that in telecentric mounts, each point of the image is illuminated by rays that have different orientations within the etalon. Therefore, appropriate rotations of the principal plane are needed over the integration domain. Since $\phi'$ is the azimuthal angle of the principal plane (Fig.~\ref{xyz}), its relation to the image plane azimuth $\phi$ can be found by geometrical considerations and depend on the ray coordinates on the pupil and chief ray position on the image plane:
\begin{equation}
\phi\prima=\tan^{-1}\left(\frac{\eta_0-r\sin\phi}{\xi_0-r\cos\phi}\right).
\end{equation}
Now, none of the factors in Eq. (\ref{exactH}) can be taken out of the integral and the expressions must be calculated numerically.
We can only find an analytical expression for the Jones matrix elements at the center of the image plane, assuming that the optical axis is perpendicular and that  $\xi_0=\eta_0=0$. Then $\phi'=\phi$ and we can simplify Eq.~(\ref{exactH}) as
\begin{equation}
\begin{gathered}
\Htildeel\prima_{11}=\int_{0}^{R_{\rm p}}\!\!\!\!\!r\H_{11}(r)\,{\rm d}r\int_{0}^{2\pi}\!\!\!\!\!\!\cos^2\phi\,{\rm d}\phi+ \int_{0}^{R_{\rm p}}\!\!\!\!\!r\H_{22}(r)\,{\rm d}r\int_{0}^{2\pi}\!\!\!\!\!\!\sin^2\phi \, {\rm d}\phi=\frac{1}{2}(\Htildeel_{11}+\Htildeel_{22}),\\
\Htildeel\prima_{22}=\int_{0}^{R_{\rm p}}\!\!\!\!\!r\H_{11}(r)\,{\rm d}r\int_{0}^{2\pi}\!\!\!\!\!\!\sin^2\phi \, {\rm d}\phi+ \int_{0}^{R_{\rm p}}\!\!\!\!\!r\H_{22}(r)\,{\rm d}r\int_{0}^{2\pi}\!\!\!\!\!\!\cos^2\phi \, {\rm d}\phi=\frac{1}{2}(\Htildeel_{11}+\Htildeel_{22}),\\
\Htildeel'_{12}=\Htildeel'_{21}=\int_{0}^{R_{\rm p}}\!\!\!\!\!r\left[\H_{11}(r)-\H_{22}(r)\right]\,{\rm d}r\int_{0}^{2\pi}\!\!\!\!\!\!\sin\phi\cos\phi \, {\rm d}\phi=0,\\
\end{gathered}
\label{eq:Htele}
\end{equation}
where $\Htildeel_{11}$ and $\Htildeel_{22}$ were defined in Eq. (\ref{Htilde}).

\section{B: Mueller matrix coefficients calculation}\label{Aa}
To calculate the coefficients $a$, $b$, $c$, and $d$ of the Mueller matrix, we follow their definitions and use the nomenclature defined in Eqs. (\ref{zeta})-(\ref{sigma}). We will employ the following definition of the Pauli matrices to be consistent with our sign convention \citep{ref:spectropolarimetry}:\footnote{Differences in the sign of the Pauli matrices lead to different conventions on the clockwise or anti-clockwise rotation of the electric field polarization. For a more detailed discussion, please visit Appendix A of \citep{ref:jefferies}}
\begin{equation}
\sigma_0=\begin{pmatrix}
1 & 0\\
0 & 1
\end{pmatrix},\:\:
\sigma_1=\begin{pmatrix}
1 & 0\\
0 & -1
\end{pmatrix},\:\:
\sigma_2=\begin{pmatrix}
0 & 1\\
1 & 0
\end{pmatrix},\:\:
\sigma_3=\begin{pmatrix}
0 & i\\
-i & 0
\end{pmatrix}.
\end{equation}
Now, according to Eq. (\ref{eq:jefferies})
\begin{equation}
a=\frac{1}{\left[1+F\sin^2\left(\cfrac{\delta_o}{2}\right)\right]\left[1+F\sin^2\left(\cfrac{\delta_o+\varphi}{2}\right)\right]}\left[\frac{\tau_o+\tau_e}{2}+\cfrac{F}{2}\tau_e\sin^2\left(\frac{\delta_o}{2}\right)+\cfrac{F}{2}\tau_o\sin^2\left(\frac{\delta_o+\varphi}{2}\right)\right]=\cfrac{\taueff}{\zeta}\left(\frac{\bar{\tau}}{\taueff}+\Gamma\right).
\end{equation}
Similarly,
\begin{equation}
b=\frac{1}{\left[1+F\sin^2\left(\cfrac{\delta_o}{2}\right)\right]\left[1+F\sin^2\left(\cfrac{\delta_o+\varphi}{2}\right)\right]}\left[\frac{\tau_o-\tau_e}{2}+\frac{F}{2}\tau_o\sin^2\left(\frac{\delta_o+\varphi}{2}\right)-\frac{F}{2}\tau_e\sin^2\left(\frac{\delta_o}{2}\right)\right]=\frac{\taueff}{\zeta}\Lambda.
\end{equation}
For $c$ and $d$ we shall first calculate $\H_{22}\H_{11}^\ast$ and $\H_{11}\H_{22}^\ast$:

\begin{equation}
\H_{22}\H_{11}^\ast=\frac{\taueff}{\zeta(1-R)^2}\left[{\rm e}^{{\rm i}\varphi/2}+R^2{\rm e}^{-{\rm i}\varphi/2}-2R\cos(\delta_o+\varphi/2)\right],
\end{equation}
\begin{equation}
\H_{11}\H_{22}^\ast=\frac{\taueff}{\zeta(1-R)^2}\left[{\rm e}^{-{\rm i}\varphi/2}+R^2{\rm e}^{{\rm i}\varphi/2}-2R\cos(\delta_o+\varphi/2)\right].
\end{equation}
Therefore, we can express $c$ and $d$ as
\begin{equation}
c=\frac{\taueff}{\zeta}\left[\frac{\cos(\varphi/2)}{(1-R)^2}+\frac{F}{4}R\cos(\varphi/2)-\frac{F}{2}\cos(\delta_o+\varphi/2)\right]=\frac{\taueff}{\zeta}\left[\frac{\cos(\varphi/2)}{(1-R)^2}+\Psi\right],
\end{equation}
\begin{equation}
d=\frac{\taueff}{\zeta}\left[\frac{-\sin(\varphi/2)}{(1-R)^2}-\frac{F}{4}R\sin(\varphi/2)\right]=\frac{\taueff}{\zeta}\left[\frac{-\sin(\varphi/2)}{(1-R)^2}+\Omega\right].
\end{equation}

\section{C: Electro-optic and piezo-electric effects in Z-cut Lithium Niobate etalons}
In $Z$-cut Lithium Niobate etalons, tuning of the transmitted wavelength is made by applying an electric field along the $Z$-cut direction. LiNbO$_3$ is an electro-optical material that presents changes in the refractive index by application of an external electric field through the Pockels effect. Changes in the width of the etalon also occur due to piezo-electric effects. Both have an influence on the birefringence of the crystal. In this Appendix, we obtain a more general expression than Eq. (\ref{varphi1}) for the birefringence  that also takes into account the presence of external fields. 

The Pockels effect depends on the particular optical axis of the crystal, but also on the direction of the incoming light and on the direction of the electric field. At an atomic level an electric field applied to certain crystals causes a redistribution of bond charges and possibly a slight deformation of the crystal lattice. In general, these alterations are not isotropic; that is, the changes vary with direction in the crystal and, therefore, the permeability tensor is no longer diagonal in presence of an external electric field \citep[e.g.,][]{ref:kasap}.

Consequently, even if the applied external field direction coincides with the optical axis ($Z$ in this case), there is no guarantee that for normal illumination no birefringence will appear. This will depend on the crystal symmetry class, which determines the form of the electro-optical tensor and not only on the direction of the incoming light and on the direction of the optical axis. For example, an uniaxial $Z$-cut crystal like KDP (KH$_2$PO$_4$) or Lithium Niobate (LiNbO$_3$), might become biaxial when applying an external field along the $Z$-axis. In the case of KDP, the field along $Z$ rotates the principal axes by 45$\degree$ about $Z$ and changes the principal indices $n_1$ and $n_2$. The particular effect of applying an electric field for Lithium Niobate need to be studied for our specific application.
\subsection{Pockels effect}
The Pockels effect consists of a linear change in the impermeability tensor  due to the linear electro-optic effect when an electric field is applied. The impermeability tensor is defined as $\imperm\equiv\epsilon_0\bepsilon^{-1}$ , where $\epsilon_0$ is the vacuum permittivity and $\bepsilon$ is the permittivity tensor. This tensor is diagonal in the principal coordinates with elements $\eta_{11}=1/n_1^2$, $\eta_{22}=1/n_2^2$, and $\eta_{33}=1/n_3^2$. The change in the impermeability tensor can be expressed as
\begin{equation}
\Delta\eta_i=\sum_{j}r_{ij}E_j,
\end{equation}
where $r_{ij}$ are the components of the electro-optical tensor and $E_j$ are the components of the electric field. Subindices $i$ and $j$ take the values $i=1,...,6$ and $j=1,2,3$. The new impermeability tensor, $\bar{\imperm}$,  is no longer diagonal in the principal dielectric axes system:
\begin{equation}
\begin{gathered}
\bar{\imperm}=
\begin{pmatrix}
1/n_1^2+\Delta\eta_1 & \Delta\eta_6 & \Delta\eta_5\\
\Delta\eta_6 & 1/n_2^2+\Delta\eta_2 & \Delta\eta_4 \\
\Delta\eta_5 & \Delta\eta_4 & 1/n_3^2+\Delta\eta_3 \\
\end{pmatrix}.
\end{gathered}
\end{equation}
The presence of cross terms
indicates that the principal dielectric axis system is
changed. Determining the new principal axes and the new refraction indices requires that the impermeability tensor is diagonalized, thus determining its eigenvalues and eigenvectors. Lithium Niobate is a trigonal 3m point group crystal \citep{ref:nikogosyan} and, therefore, its electro-optical tensor is given by \citep{ref:bass}

\begin{equation}
r=
\begin{pmatrix}
0 & -r_{22} & r_{13} \\
0 & r_{22}  & r_{13} \\
0 & 0       & r_{33} \\
0 & r_{51}  & 0      \\
r_{51} & 0  & 0 \\
-r_{22} & 0 & 0 \\
\end{pmatrix},
\end{equation}
where $r_{22}, r_{13}, r_{33}$ and $r_{51}$ depend on both the material and the specific sample. We can take the values $r_{22}\simeq 6.8, r_{13}\simeq 9.6, r_{33}\simeq 30.9$ and $r_{51}\simeq 32.6$ (all in pm/V) at $\lambda=632.8$ nm  as reference for LiNbO$_3$ \citep{ref:nikogosyan}.
If we apply an electric field along the optical axis ($j=3$):
\begin{equation}
\begin{gathered}
\Delta\eta_1=\Delta\eta_2= r_{13}V/h,\\
\Delta\eta_3= r_{33}V/h,\\
\Delta\eta_4=\Delta\eta_5=\Delta\eta_6=0,
\end{gathered}
\end{equation}
where $V$ is the associated potential difference associated to the applied electric field. In this case, the impermeability tensor is symmetric and the new refraction indices, $\bar{n}_1$, $\bar{n}_2$, $\bar{n}_3$ are given by

\begin{equation}
\bar{n}_1=\bar{n}_2\equiv \bar{n}_o(V)= \frac{n_o}{\sqrt{1+r_{13}Vn_o^2/h}} \simeq n_o-\frac{r_{13}n_o^3V}{2h},
\label{noV}
\end{equation}

\begin{equation}
\bar{n}_3(V)=\frac{n_3}{\sqrt{1+r_{33}Vn_3^2/h}} \simeq n_3-\frac{r_{33}n_3^3V}{2h}.
\label{neV}
\end{equation}
Note that Eq.~(\ref{noV}) coincides with the known unclamped Pockels effect formula for LiNbO$_3$ (Eq.~(27) of Paper I). This leads us to a explicitly modified relation between the $n_e$ and $n_o$ that takes into account both the incidence angle of the incoming light and the applied voltage employed to tune the etalon:

\begin{equation}
\frac{1}{{n}_{e}^{2} (\theta,V)}=\frac{1}{\bar{n}_{o}^{2}(V)}\cos^2\theta+\frac{1}{\bar{n}_{3}^{2}(V)}\sin^2\theta.
\label{nee2}
\end{equation}
Very interestingly, since the impermeability tensor is diagonal and $\bar{n}_1=n_2'$ for a $Z$-cut LiNbO$_3$ when an electric field in the direction of the optical axis is applied, the crystal remains uniaxial and there is no birefringence induced at normal illumination, no matter the intensity of the electric field. For $\theta\neq 0$ the birefringence is both angle and voltage dependent.

\subsection{Piezo-electric effect}
There is a second important effect that happens in LiNbO$_3$ when applying an electric field: the piezo-electric effect. It consists of a change of shape due to the application of an electric field and can be described by a linear relationship between the acting voltage and the change of width of the etalon. If the electric field is applied along the optical axis direction, the change of width is described \citep{ref:gaylor} by
\begin{equation}
\Delta h=d_{33}V,
\label{deltah}
\end{equation}
where $d_{33}\simeq 26$ pm/V \citep{ref:nikogosyan}. We can check whether the piezo-electric and electro-optical coefficients obtained from \cite{ref:nikogosyan} agree with the measured voltage tuning sensitivity found in \cite{imax}: $0.0335$ pm/V for the IMaX instrument aboard Sunrise. The estimated value is given by
\begin{equation}
\frac{\Delta\lambda}{V}=\left(d_{33}-\frac{n_o^3r_{13}}{2}\right)\frac{\lambda_0}{h}\simeq-0.078 \rm pm/V,
\end{equation}
which is twice larger than the experimental value. This departure from the measured value can be due to the fact that the electro-optical coefficients depend on the specific sample of Lithium Niobate material and on the wavelength. However, although these results differ considerably, we can use these piezo-electric and electro-optical coefficients to get a quantitative estimation of the order of magnitude of the birefringence $\varphi$. Using Eqs.~(\ref{varphi1}), (\ref{noV}), (\ref{neV}) and (\ref{deltah}), it is straightforward to show that

\begin{equation}
\varphi'(V) \simeq \frac{4\pi (h+d_{33}V)}{\lambda \cos\theta'}\left[n_o-n_3+\frac{V(r_{13}n_o^3-r_{33}n_3^3)}{2(h+d_{33}V)}\right]\sin^2(\theta'-\theta_3).
\end{equation}
Notice that $\theta'$ will also depend on $V$ as $n$ depends on the ordinary and extraordinary indices (Eq.~\ref{na}). The maximum relative variation of $\varphi(V)$ with respect to $\varphi(V=0)$ happens at the limits of the recommended range of voltages, $\pm3000$ V, of a commercial CSIRO etalon \citep{imax} and turns out to be  $\simeq 1.4\%$ if we use the above experimental electro-optical and piezo-electric coefficients \citep{ref:nikogosyan} and $n_o=2.29$, $n_e=2.20$, $R=0.92$, $A=0$, $\lambda=617.3$ nm. This variation is very small compared to the birefringence produced by other effects and has been neglected in this work.

\section{D: Exact expressions for the retardance in uniaxial media}
\label{Exact_phase}
\begin{figure}[b]
	\plottwo{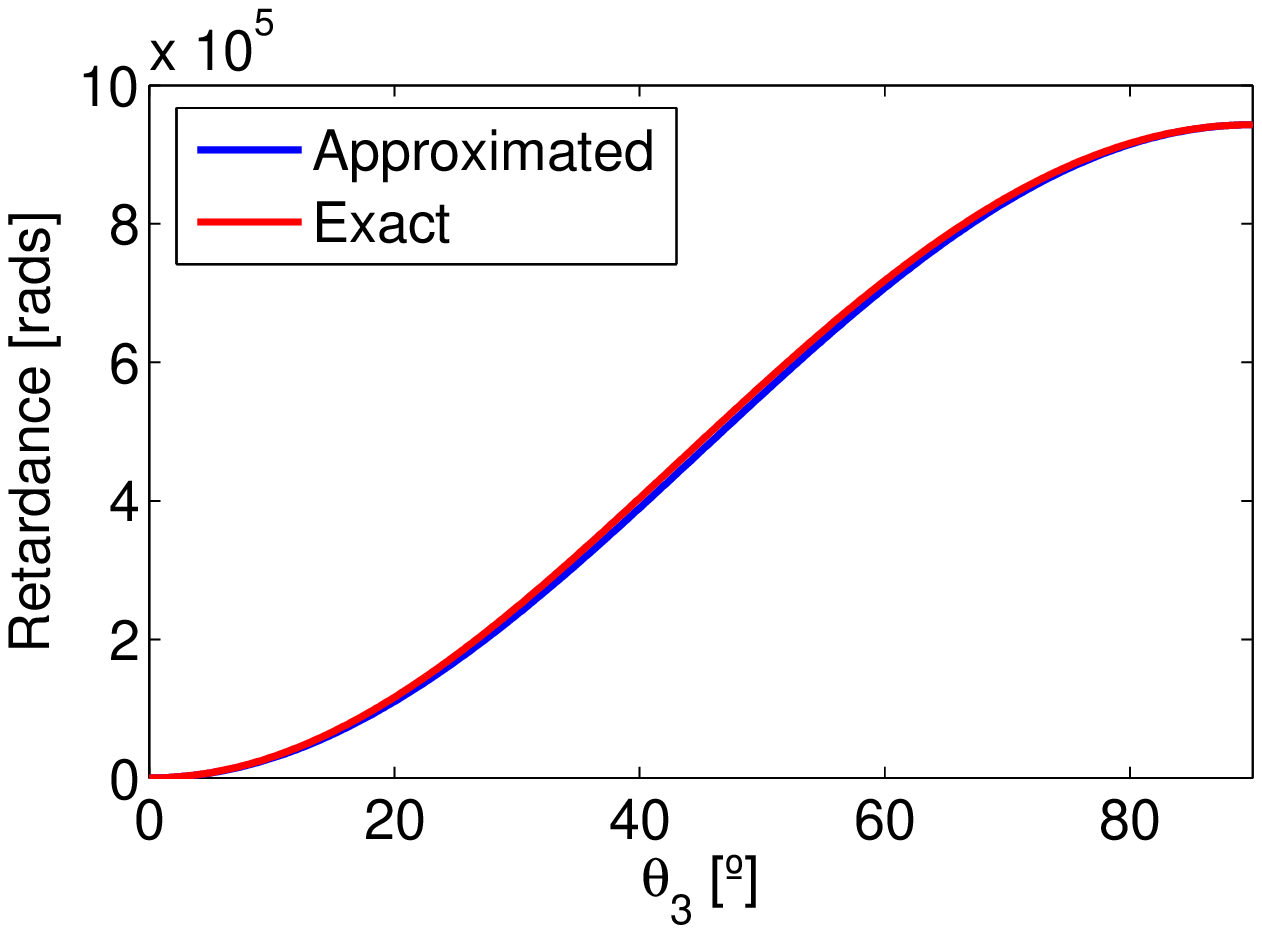}{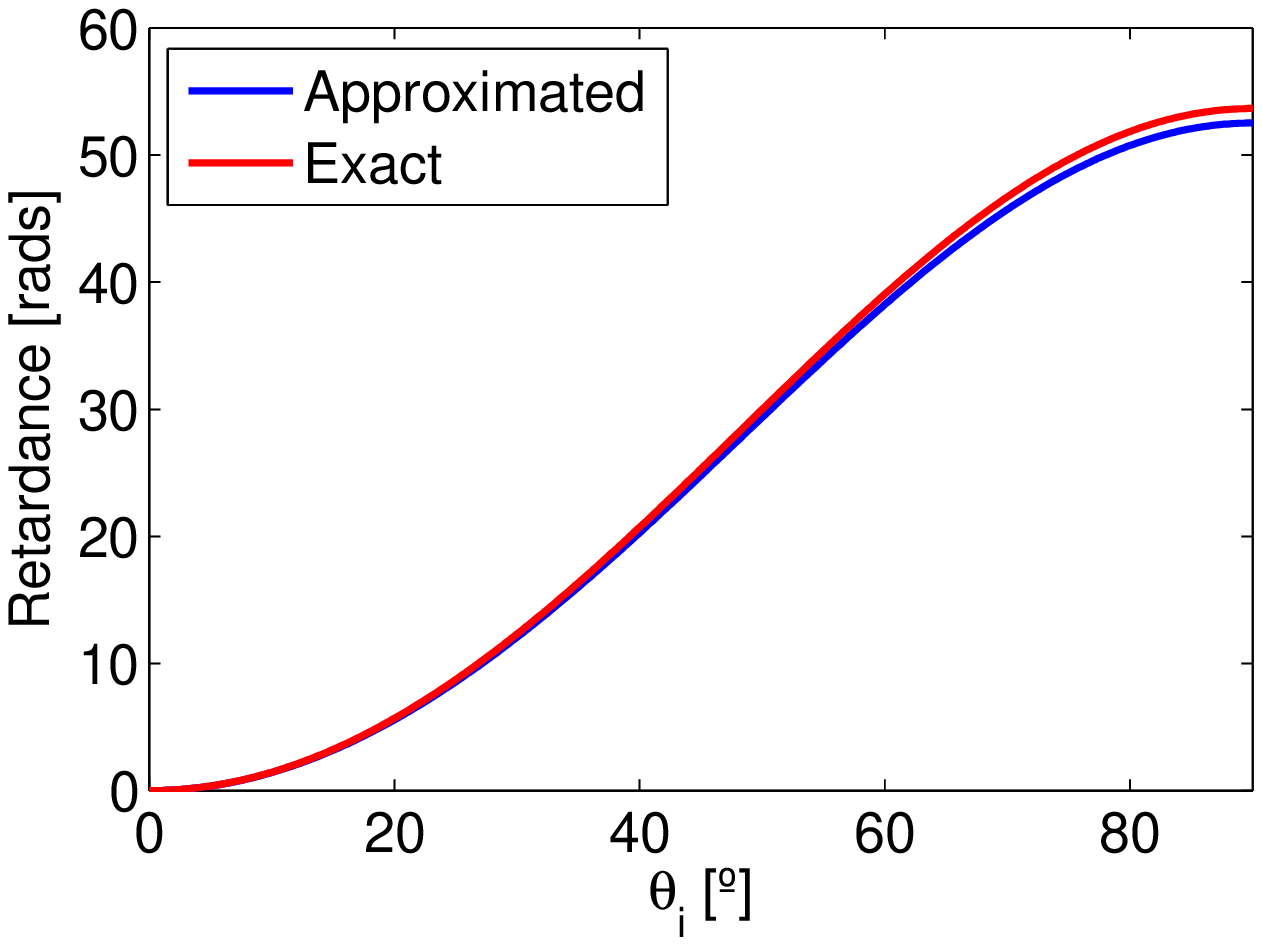}
	\caption{Retardance calculated by the approximated Equation \ref{varphi1} (blue solid line) and the exact expression found by Veiras (red solid line) as a function of the polar angle of the optical axis at normal illumination (left) and as a function of the incidence angle for an optical axis perpendicular to the etalon reflective surfaces (right).}
	\label{phi_vs_theta}
\end{figure}

A completely general calculation of phase shifts between orthogonal components of the electric field in uniaxial media was found by \cite{ref:veiras} taking into account the orientation of the optical axis for any plane wave with an arbitrary incident direction. Their results are not restricted to small birefringence media, in contrast to Equation (\ref{varphi1}). They also consider the orientation of the plane of incidence. \cite{ref:veiras} expressions and ours should be completely equivalent in the small birefringence regime. Figure \ref{phi_vs_theta} shows a comparison between \cite{ref:veiras} expressions and Eq. (\ref{varphi1}) in two particular cases, namely for normal illumination with a variable polar angle of the optical axis and for an optical axis perfectly perpendicular to the interphase with a variable incidence angle. We have employed the same parameters of the Lithium Niobate etalon used throughout this work. We can observe that differences between the exact and approximated expressions are almost negligible with $\theta_3$ for normal illumination (left) and can only be appreciated well for incidence angles higher than $50\degree$ and for $\theta_3=0$ (right).

\end{document}